\documentclass[%
reprint,
showpacs,
superscriptaddress,
amsmath,amssymb,
aps,
prb,
]{revtex4-1}

\usepackage{graphicx}
\usepackage{bm}
\usepackage{dcolumn}

\begin{document}

\title{Field-induced magnetic structures in Tb$_2$Ti$_2$O$_7$ spin liquid under field $\bm{H} \parallel [111]$}

\author{A. P. Sazonov}
\email{andrew.sazonov@frm2.tum.de}
\affiliation{Institute of Crystallography, RWTH Aachen University, D-52056 Aachen, Germany}
\affiliation{JCNS Outstation at FRM II, D-85747 Garching, Germany}
\author{A. Gukasov}
\affiliation{CEA, Centre de Saclay, DSM/IRAMIS/Laboratoire L{\'e}on Brillouin, F-91191 Gif-sur-Yvette, France}
\author{H. B. Cao}
\affiliation{Quantum Condensed Matter Division, Oak Ridge National Laboratory, Oak Ridge, Tennessee 37831, USA}
\author{P. Bonville}
\affiliation{CEA, Centre de Saclay, DSM/IRAMIS/Service de Physique de l'Etat Condens{\'e}, F-91191 Gif-Sur-Yvette, France}
\author{E. Ressouche}
\affiliation{SPSMS, UMR-E CEA/UJF-Grenoble 1, INAC, F-38054 Grenoble, France}
\author{C. Decorse}
\affiliation{ICMMO UMR 8182, Universit{\'e} Paris-Sud, F-91405 Orsay, France}
\author{I. Mirebeau}
\affiliation{CEA, Centre de Saclay, DSM/IRAMIS/Laboratoire L{\'e}on Brillouin, F-91191 Gif-sur-Yvette, France}

\date{\today}


\begin{abstract}
We have studied the field-induced magnetic structures of Tb$_2$Ti$_2$O$_7$ pyrochlore by single-crystal neutron diffraction under a field applied along the [111] local anisotropy axis, up to $H=12$\,T and down to $T=40$\,mK. We collected a hundred magnetic Bragg peaks for each field and temperature value and refined the magnetic structures with $\bm{k}=\bm{0}$ propagation vector by performing a symmetry analysis in the space group $R\bar{3}m$, reducing the number of free parameters to three only. We observe a gradual reorientation of the Tb magnetic moments towards the field direction, close to a ``3-in, 1-out~/ 1-in, 3-out'' spin structure in the whole field range 0.05--12\,T. We perform a quantitative comparison with mean-field calculations and we propose the presence of a low-temperature dynamic symmetry breaking of the local trigonal symmetry, akin to a dynamic Jahn-Teller effect, i.e. preserving the overall cubic symmetry. We discuss the possible origin of this off-diagonal mixing term in the crystal field hamiltonian in terms of quadrupole-quadrupole interaction or magneto-elastic effects.
\end{abstract}


\maketitle


\section{Introduction}

Spin ices are geometrically frustrated magnets, which support exotic ground states and excitations~\cite{sc.294.1495.2001,nt.451.42.2008}. They are observed in rare-earth pyrochlores such as $R_2$Ti$_2$O$_7$ ($R$ = Ho, Dy), where the $R$ moments are situated on a lattice of corner sharing tetrahedra, and which combine an effective ferromagnetic interaction between the $R$ moments with a local Ising anisotropy of the $R$ ion. The zero-field magnetic ground state of the spin ices has an extensive degeneracy arising from topological constraints, which also governs the positions of protons of water molecules in real ice. Recently an intensive research started studying the role of quantum fluctuations in the spin-ice regime~\cite{prb.86.075154.2012,prl.108.037202.2012}. Quantum spin-ice (QSI) or quantum spin-liquid (QSL) phases are predicted, involving tunneling between different spin or charge ice configurations. Such states could be expected in rare-earth pyrochlores under reduced local anisotropy or for low effective spin values, but none of them has been fully ascertained in a real material yet.

Tb$_2$Ti$_2$O$_7$ is a potential candidate for such states. Initial experiments performed in 1999 suggested a classical spin-liquid (SL) behavior, akin to a cooperative paramagnet where strongly correlated magnetic moments fluctuate down to 50\,mK at least~\cite{prl.82.1012.1999}. Since then, a huge number of studies has been devoted to this compound. Very recently, a transition from this SL state to mesoscopic  antiferromagnetic (AF) state with $\bm{k}=(\frac12, \frac12, \frac12)$ propagation vector was observed in powdered off-stoechiometric samples, tuned by a minute change of the Tb concentration~\cite{prb.87.060408.2013}. This static AF order coexists with the SL fluctuations since it involves a tiny ordered moment ($\sim 0.1$\,$\mu_\text{B}$). It extends over a finite correlation length, and disappears at about 0.4\,K. In single crystals, elongated diffuse spots centered at half integer Bragg peak positions were also recently observed by several groups~\cite{PhysRevLett.109.017201,prb.86.174403.2012,PhysRevB.87.094410}, using either elastic scattering or diffuse scattering of polarized neutrons. These recent observations reconcile different experimental results concerning the Tb$_2$Ti$_2$O$_7$ ground state, but their origin as well as that of the SL fluctuations remains a matter of debate, after more than 12 years of investigation.

A crucial point concerns the determination of a suitable energy interaction scheme for Tb$_2$Ti$_2$O$_7$. It is now well agreed upon that this scheme should involve, at least, a proper description of the  Tb$^{3+}$ crystal field, anisotropic first-neighbor exchange interaction between Tb moments, and dipole-dipole interactions. Starting from this basis, it was first of all suggested that unlike classical spin ices, the Tb$_2$Ti$_2$O$_7$ crystal field scheme allows some admixture of excited crystal field levels into the ground state doublet~\cite{prl.98.157204.2007,arxiv.molavian.2009}. Such an admixture may renormalize the exchange interactions of an unfrustrated antiferromagnet with local Ising anisotropy, rendering them effectively ferromagnetic, which lead the authors to call Tb$_2$Ti$_2$O$_7$ a quantum spin ice. The ground state found in this approach is however an ordered spin ice (OSI) with $\bm{k}=\bm{0}$ propagation vector, which is actually observed in Tb$_2$Sn$_2$O$_7$ (Ref.~\onlinecite{PhysRevLett.94.246402}) but not in Tb$_2$Ti$_2$O$_7$. Alternatively, a symmetry breaking of the local trigonal symmetry at the rare-earth site, inducing a two-singlet crystal field ground state for the non-Kramers Tb$^{3+}$ ion, was proposed to be the source of the SL fluctuations~\cite{prb.84.184409.2011,prb.85.054428.2012}. Such a model indeed predicts a SL phase in mean-field approximation and it accounts for the existence of a low-energy line in the inelastic neutron spectra~\cite{prb.86.174403.2012}. However, the distortion associated with this symmetry breaking, inferred from critical scattering x-ray measurements~\cite{prl.99.237202.2007}, has not been observed up to now and the static AF correlations observed below 0.4\,K cannot be understood by this model treated in mean field. Nevertheless, we believe that a part of the physical truth about the ground state in Tb$_2$Ti$_2$O$_7$ resides in the two-singlet model, and we shall use it in the following to make comparisons with the experimental data.

A way to investigate the ground state is to perturb it by a magnetic field $H$. Especially, applying the field along a local [111] anisotropy axis should provide a stringent test of the theories. Indeed, for $\bm{H} \parallel [111]$, the quantum spin-ice model predicts a magnetization plateau~\cite{arxiv.molavian.2009} as in spin ices, whereas the two-singlet model does not~\cite{arxiv.bonville.2013}. This plateau is however expected to be tiny, with anomalies in fields of about 100\,mT, occurring at very low temperatures, typically below 100\,mK. Several attempts to observe this plateau have been made recently by magnetic measurements or muon spin resonance~\cite{prl.109.047201.2012,prb.86.020410.2012,prb.86.094424.2012,PhysRevLett.110.137201}, with controversial results. No evidence of a plateau was found in the field dependence of the static magnetization, but anomalies in the ac susceptibility were detected at very low field and low temperature, well below the ``transition'' observed around 0.4\,K in zero field by specific heat and neutron scattering~\cite{prb.82.100402.2010,prb.87.060408.2013}. From the magnetization data of Ref.~\onlinecite{prl.109.047201.2012}, a phase diagram was drawn and a microscopic spin structure of the type ``all-in, all-out'' was proposed, fully different from the ``2-in, 2-out'' local structure expected for a spin ice.

In Tb$_2$Ti$_2$O$_7$, magnetic long-range order (LRO) is induced by an applied field. This LRO is clearly seen by the onset of a magnetic contribution to the Bragg peaks of the face-centered cubic crystal structure with the $\bm{k}=\bm{0}$ propagation vector, or by antiferromagnetic Bragg peaks of the simple cubic lattice with the $\bm{k}=(0,0,1)$ propagation vector. Up to now, most of the in-field neutron diffraction data~\cite{prl.96.177201.2006,prl.101.196402.2008} were obtained for $\bm{H} \parallel [110]$. In this configuration, a symmetry analysis allows one to determine the individual moment values and orientations of these complex non-collinear spin structures. This microscopic analysis emphasizes original field effects, invisible in the evolution of the average magnetization, such as a field-induced spin melting~\cite{prb.82.174406.2010} and a double-layered monopolar order~\cite{prb.85.214420.2012}. For $\bm{H} \parallel [111]$, earlier neutron data~\cite{jpcs.62.343.2001} also showed the onset of LRO but the magnetic structure was not determined.

In this work, we have performed neutron diffraction measurements in a Tb$_2$Ti$_2$O$_7$ single crystal to solve the magnetic structures for $\bm{H} \parallel [111]$ and to study the potential occurrence of a magnetization plateau. The experimental setup is described in Sec.~\ref{s:exp}. At 0.3\,K, we collected a hundred Bragg reflections for selected fields in a large field range 0.05-12\,T. We refined our data collections using symmetry analysis (Sec.~\ref{s:SymmMagn}), yielding a precise determination of the field induced magnetic structures. Our results show a continuous evolution of the moment orientations with increasing field, fully consistent with the average magnetization, and close to the so called ``3-in, 1-out~/ 1-in, 3-out'' spin structure in the whole field range ($H > 0.05$\,T) where it could be determined (Sec.~\ref{s:MagnStr}). This picture rules out the ``all-in, all-out'' structure previously proposed~\cite{prl.109.047201.2012}. Searching for a magnetization plateau, we also accurately measured the field dependence of some selected Bragg peaks at 40\,mK by scanning the magnetic field. In Sec.~\ref{s:Model}, we compare the field evolution of the magnetic structure with that predicted by a mean-field (MF) treatment, including or not the off-diagonal mixing term phenomenologically assimilated  to a dynamic distortion. We show that this term is needed to account for the data in applied field. In Sec.~\ref{s:Discussion}, we discuss our neutron results extensively with regard to the available microscopic descriptions and present some hypotheses as to the origin of the quantum mixing term, in the light of very recent experiments.


\section{\label{s:exp}Experiment}

A single crystal of Tb$_2$Ti$_2$O$_7$ was grown from a sintered rod of the same nominal composition by the floating-zone technique, using a mirror furnace~\cite{prl.101.196402.2008}. Neutron-diffraction studies were performed on the CRG-CEA diffractometer D23~($\lambda=1.2815$\,\AA) at the Institut Laue-Langevin, Grenoble, using unpolarized neutrons.

The crystal structure was characterized in zero field at 0.3\,K. A total of 256 reflections with $\ensuremath{\sin\theta/\lambda} \lesssim 0.55$\,\AA$^{-1}$ were measured and 48 unique reflections were obtained by averaging equivalent reflections ($R_\text{int} = 0.06$) using the cubic space group $Fd\bar{3}m$. These were used to refine one oxygen positional parameter, all the isotropic temperature factors, the scale factor and the extinction parameters, yielding an agreement factor $R_\text{F} = 0.05$.

To study the magnetic structures at 0.3\,K we collected about 100 Bragg reflections for each field value, in an angular range $\ensuremath{\sin\theta/\lambda} \lesssim 0.53$\,\AA$^{-1}$. The field $H$ was applied along a [111] direction and its intensity was varied between 0.05\,T and 12\,T. The program \textsc{FullProf} (Ref.~\onlinecite{phb.192.55.1993}) was used to refine the components of the Tb$^{3+}$ magnetic moments using symmetry constraints described below. We also performed field scans of selected Bragg peaks in the temperature range 40--300\,mK to study possible anomalies at very low temperatures.


\section{\label{s:SymmMagn}Magnetic symmetry analysis}

As noticed in Ref.~\onlinecite{jpcs.62.343.2001}, for $\bm{H} \parallel [111]$ we observe a magnetic contribution to the peaks of the face-centered-cubic lattice only, showing that the magnetic structure is characterized by a propagation vector $\bm{k} = \mathbf{0}$. To perform a careful analysis of the magnetic structure, we carried out a systematic search based on the theory of representations of space groups proposed by Bertaut~\cite{acra.24.217.1968} and Izyumov~\cite{jmmm.12.239.1979}. It allows one to consider all possible models of magnetic structures with a given propagation vector consistent with a crystal structure of a given space group $G$. According to this method, the magnetic structure can be expressed via the basis vectors of the irreducible representations (IRs) of the group $G$. To calculate the IRs, we used the program BasIreps from the \textsc{FullProf} suite~\cite{phb.192.55.1993}.

Applying the field along one of the four three-fold axes breaks the cubic $Fd\bar{3}m$ symmetry and leads to a magnetic structure with lower symmetry than the underlying crystal structure. Therefore none of the four IRs predicted for the space group $Fd\bar{3}m$ with $\bm{k} = \mathbf{0}$, is compatible with our data in an applied field.

We searched for new solutions in the rhombohedral space group $R\bar{3}m$, the highest subgroup of $Fd\bar{3}m$ for which a homogeneous magnetization component induced by a field applied along [111] is invariant. The assumption of a rhombohedral symmetry is \emph{a priori} valid in the paramagnetic region only (high temperature and high field) when the Zeeman energy overcomes the energy of the Tb--Tb interactions, so that a ternary symmetry of the moment values and orientations is imposed by the applied field. This point will be discussed below when comparing the results of the symmetry analysis with the mean-field calculations.

\begin{figure}
\includegraphics[width=1.0\columnwidth]{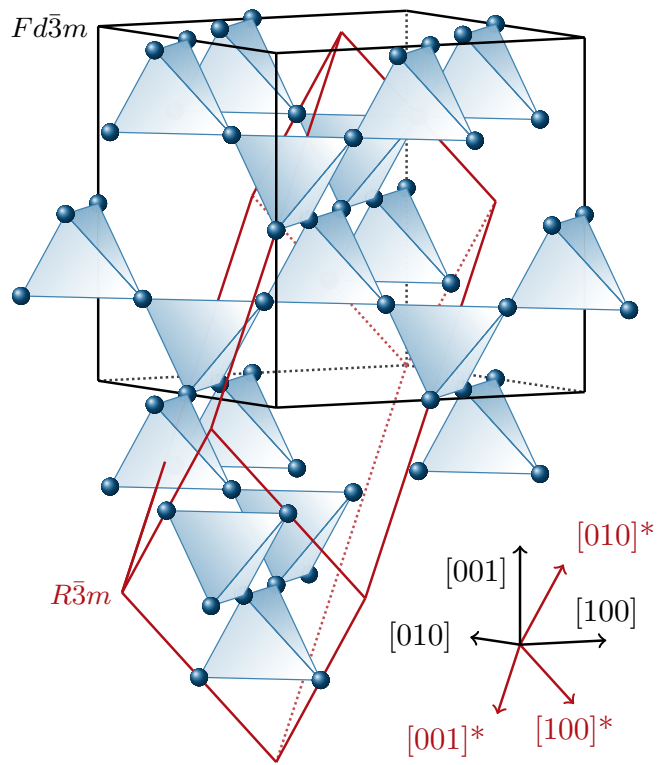}
\caption{\label{f:CrystalStructure}(Color online) Transformation from the cubic $Fd\bar{3}m$ (in black) to rhombohedral $R\bar{3}m$ (in red) space group in Tb$_2$Ti$_2$O$_7$. Only the Tb ions are shown for simplicity.}
\end{figure}

The transformation from the $Fd\bar{3}m$ space group to the $R\bar{3}m$ one is shown in Fig.~\ref{f:CrystalStructure}. The new unit-cell parameters in $R\bar{3}m$ are $a^*=b^*=a\frac{\sqrt{2}}{2}$, $c^*=a\sqrt{3}$, $\alpha^*=\beta^*=90^\circ$, and $\gamma^* = 120^\circ$, where $a$ is the cubic unit-cell parameter of the $Fd\bar{3}m$ space group. Hereafter, the symbol~* will be used to denote the lattice parameters, coordinates and directions associated with the $R\bar{3}m$ symmetry.
%
Assuming a rhombohedral symmetry leads to a splitting of the Tb $16d$ Wyckoff position in $Fd\bar{3}m$ into two crystallographically non-equivalent $3b$ and $9e$ positions in $R\bar{3}m$. The positions of the four Tb atoms within a primitive unit cell (in $R\bar{3}m$) are defined in Table~\ref{t:IrrRepTb1} as Tb1 ($3b$ position) and Tb2-4 ($9e$ position). The Tb1 site corresponds to a Tb$^{3+}$ ion whose local anisotropy axis is parallel to $\bm{H}$, whereas the Tb2-4 sites have their local anisotropy axes at 109.47\,$^\circ$ from the field.

\begin{table}
\caption{\label{t:IrrRepTb1} IR $\Gamma_3$ for the $3b$ sites and IR $\Gamma_3$ for $9e$ sites of Tb$_2$Ti$_2$O$_7$ (space group $R\bar{3}m$) associated with $\bm{k} = \mathbf{0}$. Basis vectors projected from a general vector $\bm{M}$ with the components $M_x$, $M_y$, and $M_z$. Tb1: $(0,0,0)^*$, Tb2: $(\frac{1}{6},\frac{1}{3},\frac{1}{3})^*$, Tb3: $(-\frac{1}{3},-\frac{1}{6},\frac{1}{3})^*$ and Tb4: $(\frac{1}{6},-\frac{1}{6},\frac{1}{3})^*$.}
\begin{ruledtabular}
\begin{tabular}{llrrr}
IR                 &Atom   &$M_x$    &$M_y$   &$M_z$  \\
\colrule
$\Gamma^{3b}_{3}$  &Tb1    &0~       &0~      &$U_z$  \\
$\Gamma^{9e}_{3}$  &Tb2    &$V_x$    &$2V_x$  &$V_z$  \\
            	   &Tb3    &$-2V_x$  &$-V_x$  &$V_z$  \\
                   &Tb4    &$V_x$    &$-V_x$  &$V_z$  \\
\end{tabular}
\end{ruledtabular}
\end{table}

As a result, the basis functions of the Tb1 and Tb2-4 sites belong to different IRs. Thus, the magnetic representation for the $3b$ and $9e$ sites of the SG $R\bar{3}m$ can be written as $\Gamma^{3b}_\text{Tb} = 1\,\Gamma^{(1)}_{3} + 1\,\Gamma^{(2)}_{6}$ and  $\Gamma^{9e}_\text{Tb} = 1\,\Gamma^{(1)}_{1} + 2\,\Gamma^{(1)}_{3} + 3\,\Gamma^{(2)}_{6}$, respectively. Here, the coefficients before IR denote the number of times this representation is contained in the global magnetic representation, the superscript is the dimension of IR and the subscript is the number of IR following the numbering scheme of Kovalev~\cite{book.kovalev.1965}. The basis vectors describing the Tb moments which transform according to the irreducible representations $\Gamma_6$ for the $3b$ sites and $\Gamma_1$ for $9e$ sites can be ruled out on the basis of magnetization measurements as none of them allows the existence of a net ferromagnetic moment along the field direction [001]$^*$ ([111] in cubic notation). In contrast, the one-dimensional representation $\Gamma_{3}$ for the $3b$ sites described in Table~\ref{t:IrrRepTb1} allows the presence a ferromagnetic order along the [001]$^*$ direction only (see the $M_z$ component in Table~\ref{t:IrrRepTb1}). In the case of the $9e$ sites, the one-dimensional representation $\Gamma_{3}$ (Table~\ref{t:IrrRepTb1}) allows both a net magnetization component along [001]$^*$ (see the $M_z$ component in Table~\ref{t:IrrRepTb1}) and ``antiferromagnetic'' order in the plane perpendicular to [001]$^*$ (see the $M_x$ and $M_y$ components). Thus, the Tb magnetic moments at $3b$ and $9e$ sites can be varied separately.

To summarize, by using the above symmetry constraints, one can use only one parameter $U_z$ for the Tb at $3b$ site and two other parameters $V_x$ and $V_z$ for the $9e$ sites. Therefore, the symmetry analysis reduces the number of parameters to be refined from twelve in an unconstrained refinement (three for each of the four Tb moments in a tetrahedron) to three only.


\section{\label{s:MagnStr}Magnetic structure refinement}

\begin{figure}
\includegraphics[width=1.0\columnwidth]{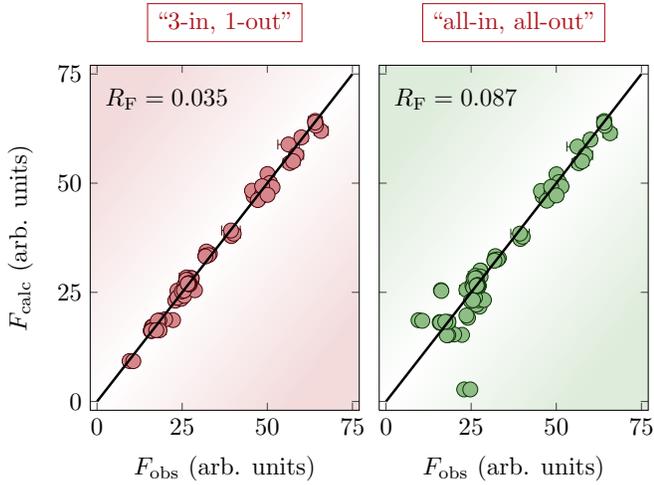}
\caption{\label{f:Refinement}(Color online) Typical examples of the agreement between the calculated, $F_\text{calc}$, and experimentally measured, $F_\text{obs}$, Bragg intensities for Tb$_2$Ti$_2$O$_7$. Experimental data were collected at $T=0.3$\,K and $H=1$\,T applied along the [111] direction. The magnetic structure refinement was performed by using the symmetry constraints described in Sec.~\ref{s:SymmMagn} within the ``3-in, 1-out~/ 1-in, 3-out'' (left panel, Fig.~\ref{f:MagnStr}) and ``all-in, all-out'' (right panel) magnetic structures, respectively.}
\end{figure}

For each field value, the integrated intensities of the collected Bragg reflections were used to refine the components of the Tb$^{3+}$ magnetic moments using the symmetry constraints described in Sec.~\ref{s:SymmMagn}. The parameters of the crystal structure were taken from the low-temperature measurements in zero field. In agreement with the above considerations, we found that among all possible irreducible representations predicted in the $R\bar{3}m$ SG for $\bm{k} = \mathbf{0}$, only $\Gamma_{3}$ (see Table~\ref{t:IrrRepTb1}) provides a reliable description of the neutron diffraction data. The agreement factors $R_\text{F}$ are about 0.04. An example of a typical fit is shown in Fig.~\ref{f:Refinement} (left panel).

\begin{figure}
\includegraphics[width=1.0\columnwidth]{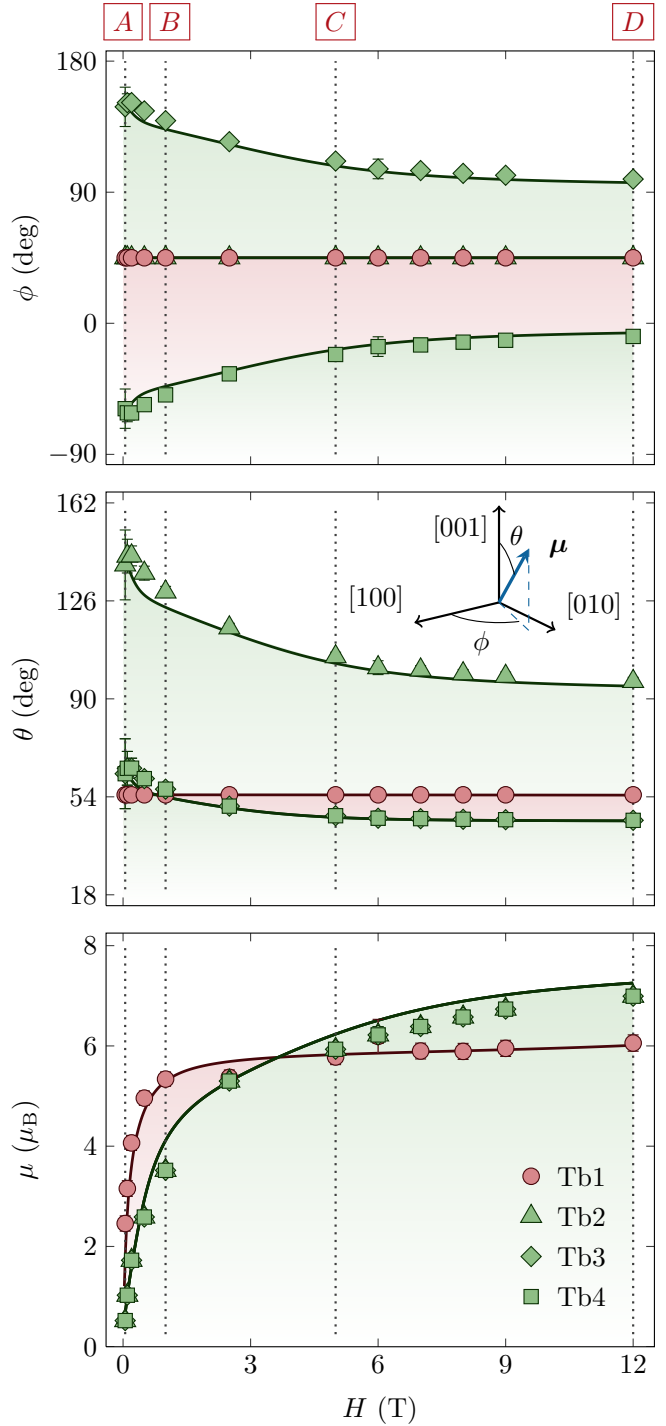}
\caption{\label{f:MagnStrParams}(Color online) Field dependence of the angles and magnitudes of the Tb magnetic moments in Tb$_2$Ti$_2$O$_7$ at 0.3\,K with $\bm{H} \parallel [111]$. Error bars are smaller than the symbol size if not given. The angles $\phi$ and $\theta$ of a given moment $\mu$ are defined at the central inset in spherical coordinates. The solid lines are calculations using the second variant of Model~II described in Sec.~\ref{s:Model}. The magnetic structures corresponding to the fields marked with ``$A$'', ``$B$'', ``$C$'' and ``$D$'' (vertical dotted lines) are shown in Figs.~\ref{f:MagnStrEvol} and~\ref{f:MagnStr} and described in the text.}
\end{figure}

\begin{figure}
\includegraphics[width=1.0\columnwidth]{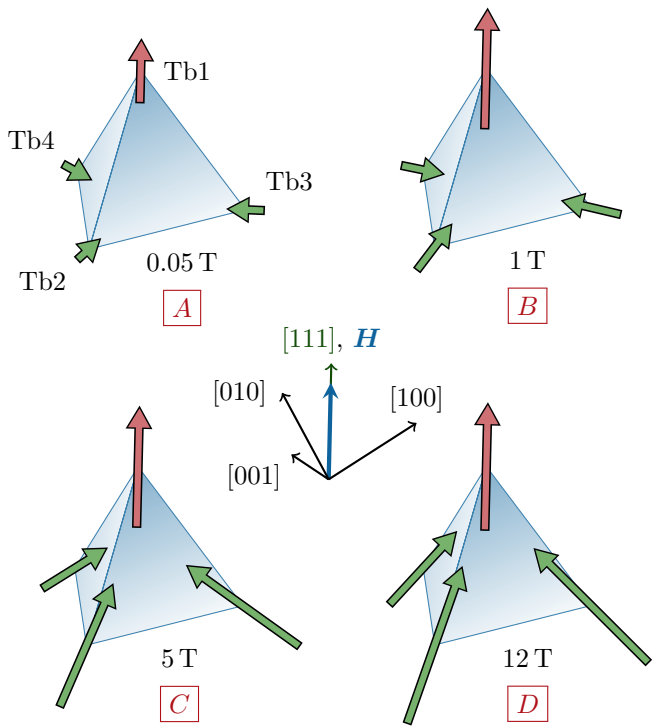}
\caption{\label{f:MagnStrEvol}(Color online) Field dependence of the Tb magnetic moments of Tb$_2$Ti$_2$O$_7$ at 0.3\,K under $\bm{H} \parallel [111]$, for typical field values (cases $A$ to $D$ in Fig.~\ref{f:MagnStrParams}). Only a single tetrahedron is shown for simplicity.}
\end{figure}

\begin{figure}
\includegraphics[width=1.0\columnwidth]{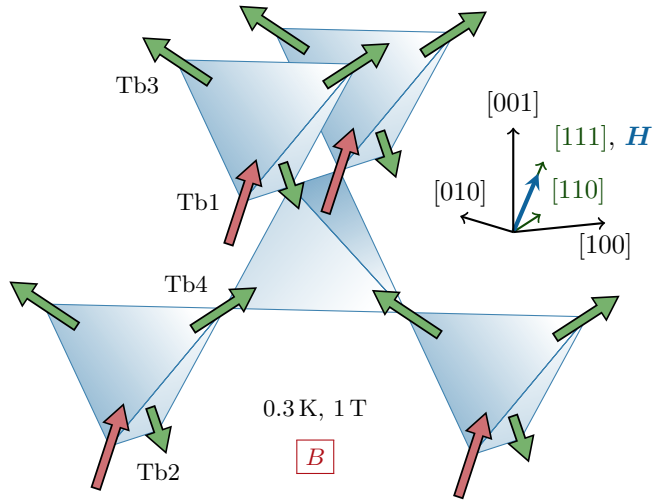}
\caption{\label{f:MagnStr}(Color online)  Field-induced magnetic structure of Tb$_2$Ti$_2$O$_7$ with $\bm{k} = \mathbf{0}$ propagation vector, stabilized for a field of 1\,T along the [111] direction (case $B$ in Fig.~\ref{f:MagnStrParams}).}
\end{figure}

The field variations of the modulus of the Tb moments, of their polar ($\theta$) and azimuthal ($\phi$) angles in the cubic frame are presented in Fig.~\ref{f:MagnStrParams}. The symbols correspond to the experimental values, determined by using the symmetry constraints (IR $\Gamma_3$, Sec.~\ref{s:SymmMagn}), whereas the solid lines refer to the mean-field calculations (Model~II, Sec.~\ref{s:Model}). The data points for the Tb1 site are in red, whereas those for the Tb2-4 sites are in green.

In Fig.~\ref{f:MagnStrEvol}, the moment orientations deduced from the refinements are drawn for different field values as the field increases from 0.05\,T to 12\,T, focusing on a single tetrahedron. Figure \ref{f:MagnStr} shows an extended plot of the magnetic lattice for $H=1$\,T. For this field value, all Tb moments are aligned close to their local $\left\langle 111 \right\rangle$ axes.

\begin{figure}
\includegraphics[width=1.0\columnwidth]{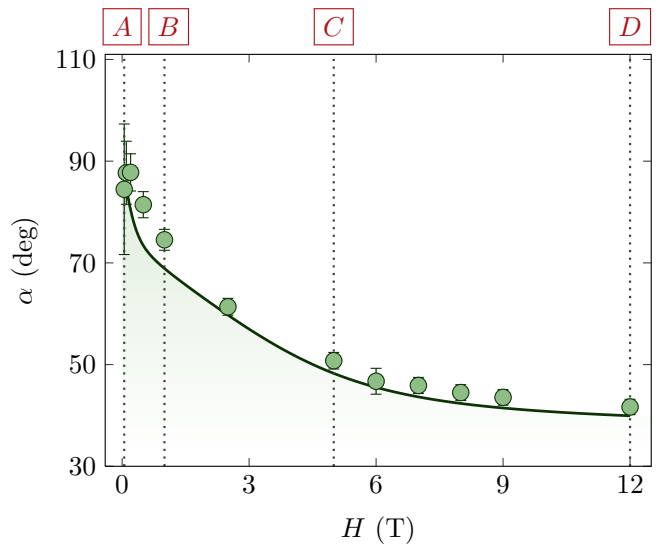}
\caption{\label{f:Angle}(Color online) Field variation of the angle $\alpha$ between the Tb2-4 magnetic moments and the applied field $\bm{H} \parallel [111]$, as deduced from the refinement of the single-crystal neutron diffraction data with symmetry analysis (see Sec.~\ref{s:SymmMagn}). The solid line is a calculation with the second variant of Model~II described in Sec.~\ref{s:Model}.} 
\end{figure}

We first notice that in the whole field range where we collected data, the so-called ``3-in, 1-out~/ 1-in, 3-out'' structure is stabilized (Fig.~\ref{f:MagnStrEvol}), with three moments (Tb2-4, in green) pointing towards the center of the tetrahedron, and one moment (Tb1, in red) pointing out, or the reverse. Namely, even at the lowest field of 0.05\,T, we can exclude the ``all-in, all-out'' antiferromagnetic structure proposed in Ref.~\onlinecite{prl.109.047201.2012}. Such a structure obeys the symmetry analysis described above with the same IR as found experimentally but with different moment orientations, yielding a much worse agreement factor (see Fig.~\ref{f:Refinement}, right panel). Actually, when the field increases, the Tb2-4 moments slowly rotate from directions close to perpendicular to the applied field towards the field direction [111]. The angle $\alpha$ between these moments and the applied field is plotted versus the field in Fig.~\ref{f:Angle}. Therefore the ferromagnetic component of the Tb2-4 moments increases for two reasons: The moments rotate and their magnitudes increase with increasing the field. At 12\,T these moments still strongly deviate, by about 42$^\circ$, from the field direction [111]. As for the Tb1 moments, they are always collinear to the field, according to the symmetry constraints described in Sec.~\ref{s:SymmMagn}.

\begin{figure}
\includegraphics[width=1.0\columnwidth]{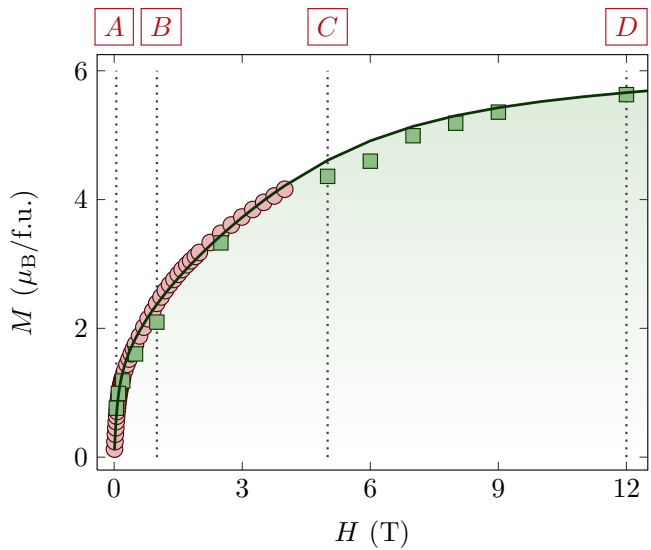}
\caption{\label{f:MvsH}(Color online) Field dependence of the magnetization $M$ of Tb$_2$Ti$_2$O$_7$ under $\bm{H} \parallel [111]$ measured at 0.08\,K according to Ref.~\onlinecite{prb.86.020410.2012} (red circles) and at 0.3\,K according to our experiments (green squares). The solid line is a calculation with Model~II described in Sec.~\ref{s:Model}.}
\end{figure}

From our experimental data, we can extract the magnetization of Tb$_2$Ti$_2$O$_7$ (green squares in Fig.~\ref{f:MvsH}) and compare it to its determination by bulk magnetic measurements~\cite{prb.86.020410.2012} (red circles in Fig.~\ref{f:MvsH}). Both quantities are in good agreement, which is an important check of the robustness and consistency of our analysis and which ensures that all the experimental corrections have been done properly. In this field range, there is almost no variation of the magnetization between 0.3\,K and 0.08\,K.

\section{\label{s:Model}Mean-field model}

\begin{figure}
\includegraphics[width=1.0\columnwidth]{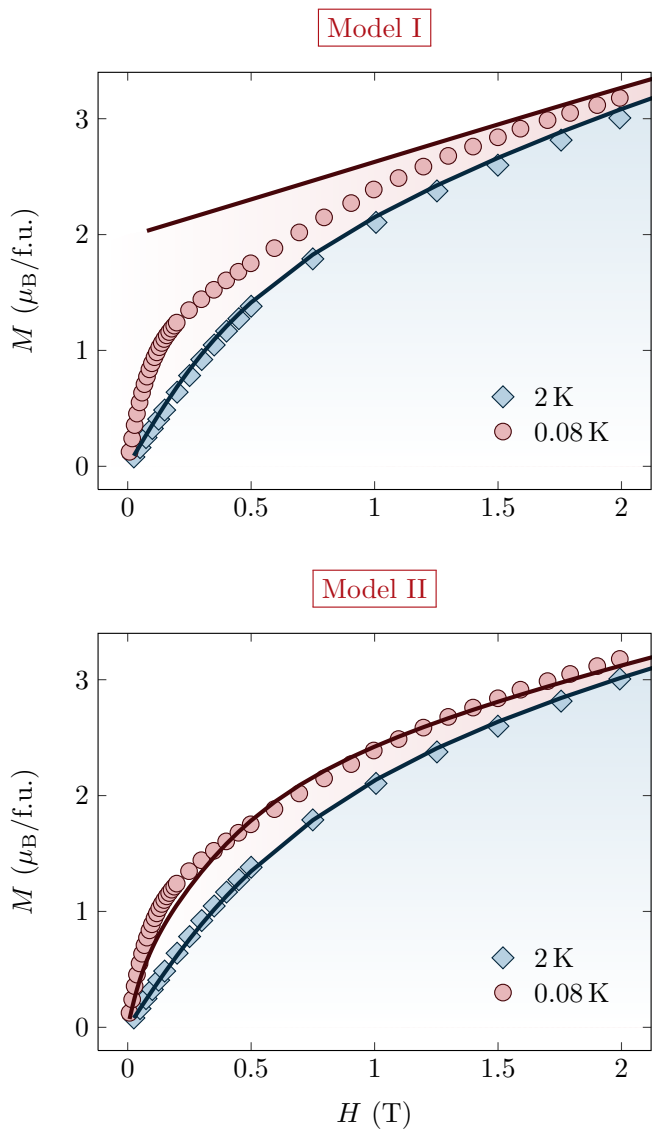}
\caption{\label{f:MeanFieldCalcM}(Color online) Low-field variations of the magnetization $M$ under $\bm{H} \parallel [111]$ measured at 0.08\,K (red circles) and 2\,K (blue diamonds) according to Ref.~\onlinecite{prb.86.020410.2012}. The solid lines are mean-field calculations without symmetry breaking (Model~I, upper panel) or with symmetry breaking (Model~II, lower panel) as described in Sec.~\ref{s:Model}.}
\end{figure}

Before computing the field evolution of the magnetic structure in Tb$_2$Ti$_2$O$_7$ for $\bm{H} \parallel [111]$, one can first try to reproduce the magnetization curves down to very low temperature. The minimal hamiltonian includes the trigonal crystal field, the Zeeman interaction and Tb--Tb interactions, i.e. anisotropic exchange and dipole-dipole coupling. The zero-field ground state is then either an ordered spin-ice or an antiferromagnetic phase, depending on parameters, with a transition temperature around 1\,K for parameter values relevant to Tb$_2$Ti$_2$O$_7$. This approach we call Model~I in the following. If one adds to the crystal field hamiltonian an off-diagonal mixing term arising at low temperature, as explained in detail in Ref.~\onlinecite{prb.84.184409.2011}, then a spin-liquid phase emerges in a certain domain of exchange parameters. This approach is called Model~II in the following. In both models, the hamiltonian is treated in a mean-field self-consistent approximation and considers the 4 Tb sites in a tetrahedron. In the frame of Model~II, an anisotropic exchange tensor $\widetilde{\cal J}_0 = \{-0.068, -0.200, -0.098\}$\,K was derived for Tb$_2$Ti$_2$O$_7$ in Ref.~\onlinecite{prb.84.184409.2011}. The low-field part of the magnetization curves at 0.08\,K and 2\,K is shown in Fig.~\ref{f:MeanFieldCalcM}, together with the curves calculated using both models and the $\widetilde{\cal J}_0$ exchange tensor. Model~I satisfactorily explains the field variation of the magnetization at 2\,K, but it strongly overestimates the low-field magnetization at the lowest temperature 0.08\,K. The ground state of Model~I with the $\widetilde{\cal J}_0$ exchange tensor is an OSI with transition temperature 0.6\,K. At 0.08\,K, the initial (vanishing field) magnetization value (2\,$\mu_\text{B}$) complies with the ice rules and is therefore worth 1/3 of the full ground state moment ($\simeq 6$\,$\mu_\text{B}$), but this remains far from the experimental data. This also confirms that the ordered spin-ice ground state stabilized in the QSI model~\cite{prl.98.157204.2007,arxiv.molavian.2009}, which predicts $\bm{k}=\bm{0}$ magnetic Bragg peaks (in zero field) as in Tb$_2$Sn$_2$O$_7$, is not the proper ground state of Tb$_2$Ti$_2$O$_7$.
 
Turning to Model~II, we first note that, as recalled in the introduction, there is presently no microscopic model able to account for all the features of the complex ground state of Tb$_2$Ti$_2$O$_7$. However, the fact that the virtual crystal field approach \cite{prl.98.157204.2007}, which implies a first order perturbation of the crystal field, cannot describe the low temperature magnetization, suggests that a zeroth order perturbation is necessary. This implies a direct quantum mixing of the wave functions of the ground doublet induced by an off-diagonal term with respect to the generic hamiltonian proposed for pyrochlores. The two-singlet model of Refs.~\onlinecite{jpcs.145.012027.2009,prb.84.184409.2011} is the simplest rationalization of a direct quantum mixing. In Tb$_2$Sn$_2$O$_7$, it accounts for the energy and $Q$ dependence of the maps measured by inelastic neutron scattering~\cite{prb.85.054428.2012} quite correctly. In Tb$_2$Ti$_2$O$_7$, although it does not account properly for the spectral weight distribution of the spin fluctuations, it provides a reasonable basis for energy integrated measurements such as diffraction or diffuse scattering~\cite{prb.84.184409.2011}. In the following, we examine the implications of Model~II for our in-field diffraction measurements.

A simple mixing term takes the form of a tetragonal distortion~\cite{prl.99.237202.2007}, with equal probabilities that its axis $OZ$ lies along one of the 3 four-fold $\left\langle 100 \right\rangle$ directions. The overall cubic symmetry is therefore preserved. In case $OZ$ is along [001], the additional mixing term is:
\begin{equation}
\label{eq:hq}
  \mathcal{H}_Q = D_Q J^2_Z,
\end{equation}
which writes in the local frame with [111] as $z$ axis,
\begin{equation}
\label{eq:hqd}
  \mathcal{H}_Q = \frac{D_Q}{3} \, \left[ 2J^2_x + J^2_z + \sqrt{2}(J_xJ_z + J_zJ_x) \right].
\end{equation}
Since a four-fold axis makes the same angle with each of the local $\left\langle 111 \right\rangle$ axes, hamiltonian ${\cal H}_Q$ has the same expression in the 4 local frames of a tetrahedron. Inspection shows that the zeroth order degeneracy lifting comes mainly from the last term in Eq.~\ref{eq:hqd}, i.e. the $J_xJ_z + J_zJ_x$ operator. The value of the parameter $D_Q$ was estimated to be 0.25\,K from high energy-resolution inelastic neutron scattering in zero magnetic field~\cite{prl.96.177201.2006,prb.84.184409.2011}. The magnetization computed using Model~II with ${\cal J}_0$ exchange and $D_Q=0.25$\,K, satisfactorily captures the low-field part (Fig.~\ref{f:MeanFieldCalcM}) as well as the whole magnetization curve up to 12\,T (Fig.~\ref{f:MvsH}) in Tb$_2$Ti$_2$O$_7$ at very low temperature, thus allowing to try and reproduce, on a more microscopic basis, the in-field magnetic structure.

\begin{figure*}
\includegraphics[width=0.98\textwidth]{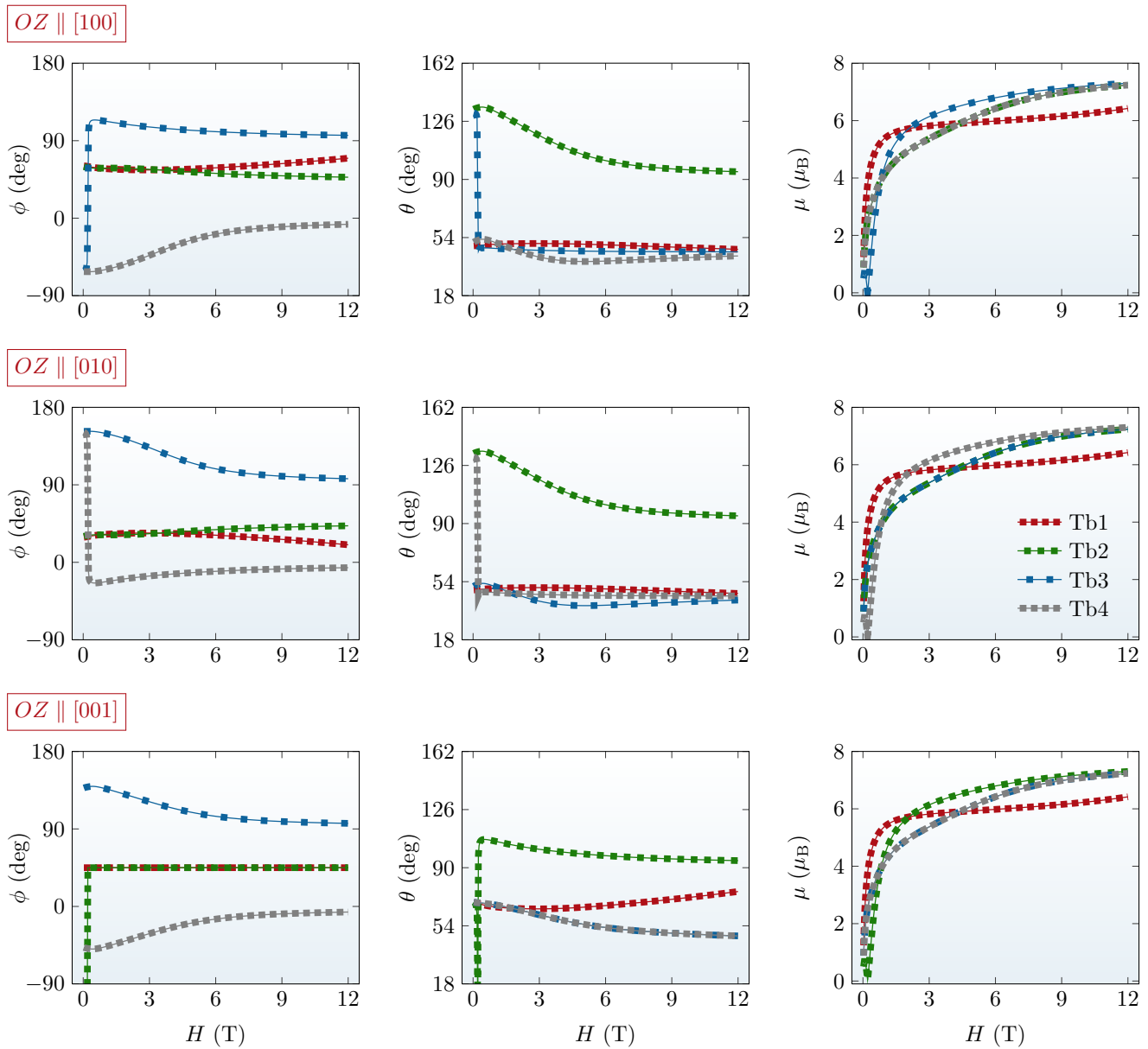}
\caption{\label{f:MeanFieldDist}(Color online) Calculated field variations, using Model~II, of the modulus of the Tb moment, of its polar ($\theta$) and azimuthal ($\phi$) angles in the cubic frame for the 4 Tb sites, and for the three possible distortion axes [100], [010] and [001].}
\end{figure*}

The calculated spin structure using Model~II depends on the particular axis $OZ$ chosen for the mixing term. The predictions for the four Tb sites in terms of moment and angle values are shown in Fig.~\ref{f:MeanFieldDist} independently for the 3 axes [001], [010] and [100]. The red curves correspond to the moment on the Tb1 site which has its three-fold axis along $\bm{H} \parallel [111]$. The magnitude of the Tb1 moment quickly saturates above about 1\,T and then does not significantly change up to 12\,T, where the experimental value $\mu_\text{Tb1} = 6.1(2)$\,$\mu_\text{B}$ is in good agreement with the calculated one. The calculated $\mu(H)$ curves for the Tb1 moment are independent of the choice of the $OZ$ axis, but $\theta(H)$ and $\phi(H)$ are slightly different, especially at high field. By contrast, both moment values and orientations on the Tb2-4 sites depend on the choice of the axis. Changing the axis from [001] to [010] or [100] induces permutations of the behaviors on the Tb2-4 sites, in such a way that in each case two of  three sites are swapped. In the following, we compare  the predictions of Model~II with the results of our neutron experiment in more detail.
 
\section{\label{s:Discussion}Discussion}

The above analysis enlightens two points. First, a ``3-in, 1-out~/ 1-in, 3-out'' magnetic structure is stabilized in the broad field range 0.05-12\,T where we performed the data collections, with a gradual evolution of the moment values and orientations. Second, an energy term directly mixing the wave-functions of the ground Tb$^{3+}$ doublet, such as proposed by Model~II, is necessary to account for the magnetization curve at low temperature. Considering the mixing hamiltonian in terms of Jahn-Teller effect, an equally probable distribution of the axis direction may be realized either at a microscopic or at a macroscopic level. Here we consider two variants, akin to the ``single-K'' (multi-domain) and ``multi-K'' (single-domain) spin structures usually considered for a non-zero propagation vector. These variants can be associated with a static and a dynamic Jahn-Teller effect, respectively. We discuss here the interpretation of the neutron results with these two variants.

In the first variant (static Jahn-Teller effect), we assume that the Jahn-Teller distortion is static, extends over a finite domain size, and that the domain orientations are equally distributed among the three cubic four-fold axes [100], [010] and [001]. The (unknown) domain size must be larger than the correlation length deduced from the width of the Bragg peaks (a few hundreds of~\AA). In this case, for each Bragg peak, one must average the contribution to the magnetic intensity from the three domains, and compare these averages to the measured Bragg intensities. This comparison does not require any constraint on the data treatment since the symmetry analysis and even the data refinement are not necessary.

In the second variant (dynamical Jahn-Teller effect), we assume that the distortion remains long-range, but that its axis changes with time on each site between the three possible orientations. In this case, one should average the amplitude of the magnetic structure factor associated with the three possible axes on each Tb site. This corresponds to a vector sum of the magnetic moments calculated for each axis direction. Here, one can compare not only the Bragg intensities, but also the moment values and angles, derived from the refinement and the symmetry analysis, to the calculated values.

\begin{figure}
\includegraphics[width=1.0\columnwidth]{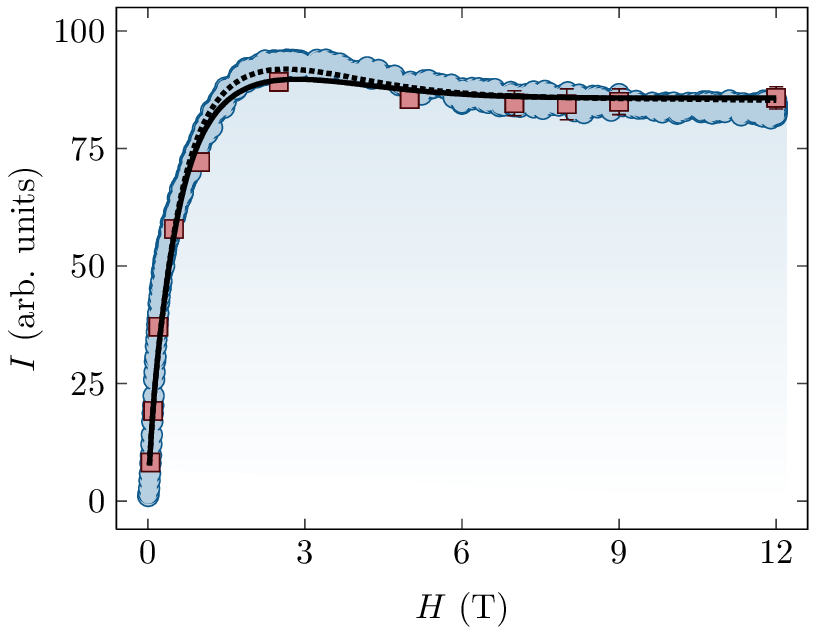}
\caption{\label{f:IvsH}(Color online) Field dependence of the ($2\bar{2}0$) Bragg reflection under $\bm{H} \parallel [111]$. The red squares correspond to the integrated intensities obtained with the data collections at 300\,mK, the blue circles are peak intensities measured during field scans at 150--200\,mK. The lines are mean-field calculations with the Model~II described in Sec.~\ref{s:Model}. The two variants of the model (solid and dashed lines) yield the same result within less that 5\,\% for all calculated Bragg peaks.}
\end{figure}

Calculations of the Bragg intensities show that the two variants never differ more than 5\,\%, and are in excellent agreement with the experimental data in the whole field range. As an example, the field dependence of the ($2\bar{2}0$) Bragg reflection is shown in Fig.~\ref{f:IvsH}. Furthermore, the moment values and angles calculated assuming the second variant also agree very well with those deduced from the refinement, as shown in Figs.~\ref{f:MagnStrParams} and~\ref{f:Angle}.

We conclude that the spin structures predicted assuming either static or dynamic Jahn-Teller effect are equally compatible with our diffraction data in applied field. A dynamic Jahn-Teller effect could explain why no distortion has been observed on a macroscopic scale up to now. The main drawback of the two-singlet model, appearing in zero field, remains however present. Namely, the mechanism slowing down the spin fluctuations (or softening the excitations) and yielding a dominant ``elastic'' signal in the neutron maps is unexplained. Up to now, no model succeeded to account for them, so that one should call for more complex mechanisms yielding cooperative effects. Their nature could be either intrinsic (quantum mixing on a larger scale than a single site yielding a Jahn-Teller transition or quadrupolar ordering) or extrinsic (influence of disorder and further neighbor interactions, yielding a spin glass). Such mechanisms could restore a magnetic state in zero field instead of the non-magnetic one found by the mean-field two-singlet model.

\begin{figure}
\includegraphics[width=1.0\columnwidth]{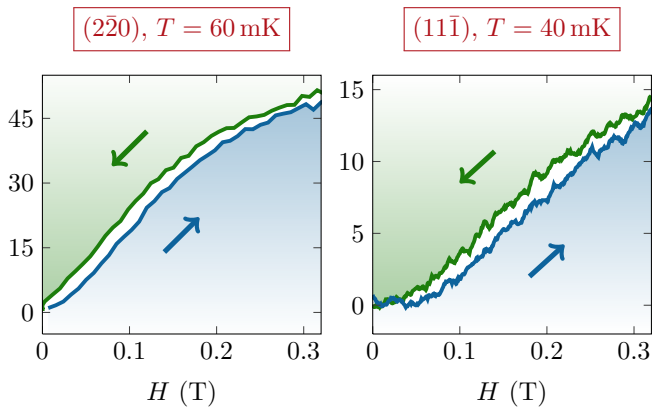}
\caption{\label{f:Hysteresis}(Color online) Field dependences of the intensity of ($2\bar{2}0$) and ($11\bar{1}$) Bragg reflections in Tb$_2$Ti$_2$O$_7$ with $\bm{H} \parallel [111]$ measured during field scans at 40--60\,mK.}
\end{figure}

We also show in Fig.~\ref{f:Hysteresis} the low-field variation of the peak intensities of some typical Bragg reflections, which we measured down to 0.04\,K by scanning the field with great accuracy, searching for anomalies related to a potential magnetization plateau. We did not observe any anomalies in the field range predicted~\cite{arxiv.molavian.2009}, and the field variations of the Bragg peaks almost superimpose in the temperature range 40--300\,mK. We therefore conclude that a magnetization plateau is quite unlikely. We however notice a hysteresis when cooling the sample in a magnetic field, which we did not investigate in detail, but which reminds of the irreversibilities~\cite{prl.109.047201.2012,prb.86.020410.2012,prb.86.094424.2012,PhysRevLett.110.153201} of the susceptibility and $\mu$SR below 0.4\,K.

Finally, we discuss possible origins for the zeroth order quantum mixing which seems necessary in order to account for many low temperature properties of Tb$_2$Ti$_2$O$_7$. In keeping with the Jahn-Teller picture, a two-ion quadrupole-quadrupole interaction naturally emerges from hamiltonian in Eq.~\ref{eq:hqd} since it can be written in terms of the rare-earth quadrupolar operators $Q_{ij}$, with their standard definitions, as:
\begin{equation}
\label{eq:hqq}
\mathcal{H}_Q = \frac{D_Q}{3} \, \left[ 2Q_{xx} + Q_{zz} + \frac{J(J+1)}{3} + 2\sqrt{2}Q_{xz} \right].
\end{equation}
As mentioned in Sec.~\ref{s:Model}, the leading term yielding a zeroth order splitting of the Tb$^{3+}$ ground doublet is $Q_{xz}=\frac12 (J_xJ_z+J_zJ_x)$. Then, keeping only this term, hamiltonian in Eq.~\ref{eq:hqq} can be seen as a Jahn-Teller driven quadrupole-quadrupole coupling treated in mean field, and it would be equivalent to:
\begin{equation}
\label{eq:hjt}
\mathcal{H}_{JT} = -g_Q \left\langle Q_{xz} \right\rangle Q_{xz},
\end{equation}
where $g_Q$ is an interaction parameter. Such an energy term should lead to a quadrupolar transition, accompanied by a lack of standard long-range magnetic order, at a temperature $T_Q$ depending on the interaction strength $g_Q$. The specific heat peak observed near 0.5\,K for a certain degree of off-stoichiometry in Tb$_2$Ti$_2$O$_7$ (Ref.~\onlinecite{prb.87.060408.2013}) could correspond to such a transition.

Another possibility is that the state mixing would take its origin in the single ion magneto-elastic coupling, known to be strong in Tb$_2$Ti$_2$O$_7$ (Refs.~\onlinecite{mamsurova,nakanishi,luan}). The magneto-elastic hamiltonian at the rare-earth site with $D_{3d}$ symmetry, limited to second order terms, also involves the $Q_{xz}$ operator (usually called $O_2^1$). Using the notations of Ref.~\onlinecite{jpcs.324.012036.2011}, the relevant part of the magneto-elastic interaction writes as:
\begin{equation}
\label{eq:hmel}
\mathcal{H}_{m-el} = \left[ (\epsilon_{xx}-\epsilon_{yy}) B_{21}^{xx} + 2\epsilon_{xz} B_{21}^{zx} \right] Q_{xz},
\end{equation}
involving components of the strain tensor $\tilde \epsilon$. A resonant coupling of a phonon to the $4f$ electrons can then occur through this interaction. The existence of such a coupling is suggested by very recent inelastic neutron scattering data showing a strong interaction between the acoustic phonon branch and the first crystal field excitation, leading to a hybrid vibronic mode~\cite{arxiv.guitteny.2013,arxiv.fennell.2013}. Very recent measurements of the thermal conductivity in Tb$_2$Ti$_2$O$_7$ concluded along the same line, i.e. that the acoustic phonon branch is strongly scattered at low temperature, resulting in a very low thermal conductivity~\cite{arxiv.li.2013}.


\section{Conclusion}

In conclusion, our single-crystal neutron diffraction data obtained in Tb$_2$Ti$_2$O$_7$ with $\bm{H} \parallel [111]$ show a continuous evolution of the field-induced magnetic structure, which keeps a ``3-in, 1-out~/ 1-in, 3-out'' arrangement in the whole field range up to 12\,T; this invalidates both the magnetization plateau and the ``all-in, all-out'' spin structures previously proposed. Comparison with a mean-field model strongly supports the existence of a zeroth order quantum mixing of the two states of the ground Tb$^{3+}$ doublet, already inferred from inelastic neutron scattering, and occurring at low temperature. The evolution of the magnetic structure deduced from our diffraction data and of the magnetization down to 0.3\,K, in the field range 0.05-12\,T, for $\bm{H} \parallel [111]$, are well accounted for by our model. Considering the quantum mixing in terms of Jahn-Teller effect, the diffraction data are compatible with both a static and a dynamic effect. Other phenomena driven by the magnetoelastic interaction, such as quadrupolar ordering and/or hybrid electronic-phonon modes as recently observed, could possibly be simulated by the phenomenological quantum mixing presented here.


\begin{acknowledgments}
We thank S. Petit and J. Robert for useful discussions, and B. Grenier for her help during the experiment.
\end{acknowledgments}



\begin{thebibliography}{39}%
\makeatletter
\providecommand \@ifxundefined [1]{%
 \@ifx{#1\undefined}
}%
\providecommand \@ifnum [1]{%
 \ifnum #1\expandafter \@firstoftwo
 \else \expandafter \@secondoftwo
 \fi
}%
\providecommand \@ifx [1]{%
 \ifx #1\expandafter \@firstoftwo
 \else \expandafter \@secondoftwo
 \fi
}%
\providecommand \natexlab [1]{#1}%
\providecommand \enquote  [1]{``#1''}%
\providecommand \bibnamefont  [1]{#1}%
\providecommand \bibfnamefont [1]{#1}%
\providecommand \citenamefont [1]{#1}%
\providecommand \href@noop [0]{\@secondoftwo}%
\providecommand \href [0]{\begingroup \@sanitize@url \@href}%
\providecommand \@href[1]{\@@startlink{#1}\@@href}%
\providecommand \@@href[1]{\endgroup#1\@@endlink}%
\providecommand \@sanitize@url [0]{\catcode `\\12\catcode `\$12\catcode
  `\&12\catcode `\#12\catcode `\^12\catcode `\_12\catcode `\%12\relax}%
\providecommand \@@startlink[1]{}%
\providecommand \@@endlink[0]{}%
\providecommand \url  [0]{\begingroup\@sanitize@url \@url }%
\providecommand \@url [1]{\endgroup\@href {#1}{\urlprefix }}%
\providecommand \urlprefix  [0]{URL }%
\providecommand \Eprint [0]{\href }%
\providecommand \doibase [0]{http://dx.doi.org/}%
\providecommand \selectlanguage [0]{\@gobble}%
\providecommand \bibinfo  [0]{\@secondoftwo}%
\providecommand \bibfield  [0]{\@secondoftwo}%
\providecommand \translation [1]{[#1]}%
\providecommand \BibitemOpen [0]{}%
\providecommand \bibitemStop [0]{}%
\providecommand \bibitemNoStop [0]{.\EOS\space}%
\providecommand \EOS [0]{\spacefactor3000\relax}%
\providecommand \BibitemShut  [1]{\csname bibitem#1\endcsname}%
\let\auto@bib@innerbib\@empty
\bibitem [{\citenamefont {Bramwell}\ and\ \citenamefont
  {Gingras}(2001)}]{sc.294.1495.2001}%
  \BibitemOpen
  \bibfield  {author} {\bibinfo {author} {\bibfnamefont {S.~T.}\ \bibnamefont
  {Bramwell}}\ and\ \bibinfo {author} {\bibfnamefont {M.~J.~P.}\ \bibnamefont
  {Gingras}},\ }\href@noop {} {\bibfield  {journal} {\bibinfo  {journal}
  {Science}\ }\textbf {\bibinfo {volume} {294}},\ \bibinfo {pages} {1495}
  (\bibinfo {year} {2001})}\BibitemShut {NoStop}%
\bibitem [{\citenamefont {Castelnovo}\ \emph {et~al.}(2008)\citenamefont
  {Castelnovo}, \citenamefont {Moessner},\ and\ \citenamefont
  {Sondhi}}]{nt.451.42.2008}%
  \BibitemOpen
  \bibfield  {author} {\bibinfo {author} {\bibfnamefont {C.}~\bibnamefont
  {Castelnovo}}, \bibinfo {author} {\bibfnamefont {R.}~\bibnamefont
  {Moessner}}, \ and\ \bibinfo {author} {\bibfnamefont {S.~L.}\ \bibnamefont
  {Sondhi}},\ }\href@noop {} {\bibfield  {journal} {\bibinfo  {journal}
  {Nature}\ }\textbf {\bibinfo {volume} {451}},\ \bibinfo {pages} {42}
  (\bibinfo {year} {2008})}\BibitemShut {NoStop}%
\bibitem [{\citenamefont {Benton}\ \emph {et~al.}(2012)\citenamefont {Benton},
  \citenamefont {Sikora},\ and\ \citenamefont {Shannon}}]{prb.86.075154.2012}%
  \BibitemOpen
  \bibfield  {author} {\bibinfo {author} {\bibfnamefont {O.}~\bibnamefont
  {Benton}}, \bibinfo {author} {\bibfnamefont {O.}~\bibnamefont {Sikora}}, \
  and\ \bibinfo {author} {\bibfnamefont {N.}~\bibnamefont {Shannon}},\ }\href
  {\doibase 10.1103/PhysRevB.86.075154} {\bibfield  {journal} {\bibinfo
  {journal} {Phys. Rev. B}\ }\textbf {\bibinfo {volume} {86}},\ \bibinfo
  {pages} {075154} (\bibinfo {year} {2012})}\BibitemShut {NoStop}%
\bibitem [{\citenamefont {Savary}\ and\ \citenamefont
  {Balents}(2012)}]{prl.108.037202.2012}%
  \BibitemOpen
  \bibfield  {author} {\bibinfo {author} {\bibfnamefont {L.}~\bibnamefont
  {Savary}}\ and\ \bibinfo {author} {\bibfnamefont {L.}~\bibnamefont
  {Balents}},\ }\href {\doibase 10.1103/PhysRevLett.108.037202} {\bibfield
  {journal} {\bibinfo  {journal} {Phys. Rev. Lett.}\ }\textbf {\bibinfo
  {volume} {108}},\ \bibinfo {pages} {037202} (\bibinfo {year}
  {2012})}\BibitemShut {NoStop}%
\bibitem [{\citenamefont {Gardner}\ \emph {et~al.}(1999)\citenamefont
  {Gardner}, \citenamefont {Dunsiger}, \citenamefont {Gaulin}, \citenamefont
  {Gingras}, \citenamefont {Greedan}, \citenamefont {Kiefl}, \citenamefont
  {Lumsden}, \citenamefont {MacFarlane}, \citenamefont {Raju}, \citenamefont
  {Sonier}, \citenamefont {Swainson},\ and\ \citenamefont
  {Tun}}]{prl.82.1012.1999}%
  \BibitemOpen
  \bibfield  {author} {\bibinfo {author} {\bibfnamefont {J.~S.}\ \bibnamefont
  {Gardner}}, \bibinfo {author} {\bibfnamefont {S.~R.}\ \bibnamefont
  {Dunsiger}}, \bibinfo {author} {\bibfnamefont {B.~D.}\ \bibnamefont
  {Gaulin}}, \bibinfo {author} {\bibfnamefont {M.~J.~P.}\ \bibnamefont
  {Gingras}}, \bibinfo {author} {\bibfnamefont {J.~E.}\ \bibnamefont
  {Greedan}}, \bibinfo {author} {\bibfnamefont {R.~F.}\ \bibnamefont {Kiefl}},
  \bibinfo {author} {\bibfnamefont {M.~D.}\ \bibnamefont {Lumsden}}, \bibinfo
  {author} {\bibfnamefont {W.~A.}\ \bibnamefont {MacFarlane}}, \bibinfo
  {author} {\bibfnamefont {N.~P.}\ \bibnamefont {Raju}}, \bibinfo {author}
  {\bibfnamefont {J.~E.}\ \bibnamefont {Sonier}}, \bibinfo {author}
  {\bibfnamefont {I.}~\bibnamefont {Swainson}}, \ and\ \bibinfo {author}
  {\bibfnamefont {Z.}~\bibnamefont {Tun}},\ }\href@noop {} {\bibfield
  {journal} {\bibinfo  {journal} {Phys. Rev. Lett.}\ }\textbf {\bibinfo
  {volume} {82}},\ \bibinfo {pages} {1012} (\bibinfo {year}
  {1999})}\BibitemShut {NoStop}%
\bibitem [{\citenamefont {Taniguchi}\ \emph {et~al.}(2013)\citenamefont
  {Taniguchi}, \citenamefont {Kadowaki}, \citenamefont {Takatsu}, \citenamefont
  {F\aa{}k}, \citenamefont {Ollivier}, \citenamefont {Yamazaki}, \citenamefont
  {Sato}, \citenamefont {Yoshizawa}, \citenamefont {Shimura}, \citenamefont
  {Sakakibara}, \citenamefont {Hong}, \citenamefont {Goto}, \citenamefont
  {Yaraskavitch},\ and\ \citenamefont {Kycia}}]{prb.87.060408.2013}%
  \BibitemOpen
  \bibfield  {author} {\bibinfo {author} {\bibfnamefont {T.}~\bibnamefont
  {Taniguchi}}, \bibinfo {author} {\bibfnamefont {H.}~\bibnamefont {Kadowaki}},
  \bibinfo {author} {\bibfnamefont {H.}~\bibnamefont {Takatsu}}, \bibinfo
  {author} {\bibfnamefont {B.}~\bibnamefont {F\aa{}k}}, \bibinfo {author}
  {\bibfnamefont {J.}~\bibnamefont {Ollivier}}, \bibinfo {author}
  {\bibfnamefont {T.}~\bibnamefont {Yamazaki}}, \bibinfo {author}
  {\bibfnamefont {T.~J.}\ \bibnamefont {Sato}}, \bibinfo {author}
  {\bibfnamefont {H.}~\bibnamefont {Yoshizawa}}, \bibinfo {author}
  {\bibfnamefont {Y.}~\bibnamefont {Shimura}}, \bibinfo {author} {\bibfnamefont
  {T.}~\bibnamefont {Sakakibara}}, \bibinfo {author} {\bibfnamefont
  {T.}~\bibnamefont {Hong}}, \bibinfo {author} {\bibfnamefont {K.}~\bibnamefont
  {Goto}}, \bibinfo {author} {\bibfnamefont {L.~R.}\ \bibnamefont
  {Yaraskavitch}}, \ and\ \bibinfo {author} {\bibfnamefont {J.~B.}\
  \bibnamefont {Kycia}},\ }\href {\doibase 10.1103/PhysRevB.87.060408}
  {\bibfield  {journal} {\bibinfo  {journal} {Phys. Rev. B}\ }\textbf {\bibinfo
  {volume} {87}},\ \bibinfo {pages} {060408} (\bibinfo {year}
  {2013})}\BibitemShut {NoStop}%
\bibitem [{\citenamefont {Fennell}\ \emph {et~al.}(2012)\citenamefont
  {Fennell}, \citenamefont {Kenzelmann}, \citenamefont {Roessli}, \citenamefont
  {Haas},\ and\ \citenamefont {Cava}}]{PhysRevLett.109.017201}%
  \BibitemOpen
  \bibfield  {author} {\bibinfo {author} {\bibfnamefont {T.}~\bibnamefont
  {Fennell}}, \bibinfo {author} {\bibfnamefont {M.}~\bibnamefont {Kenzelmann}},
  \bibinfo {author} {\bibfnamefont {B.}~\bibnamefont {Roessli}}, \bibinfo
  {author} {\bibfnamefont {M.~K.}\ \bibnamefont {Haas}}, \ and\ \bibinfo
  {author} {\bibfnamefont {R.~J.}\ \bibnamefont {Cava}},\ }\href {\doibase
  10.1103/PhysRevLett.109.017201} {\bibfield  {journal} {\bibinfo  {journal}
  {Phys. Rev. Lett.}\ }\textbf {\bibinfo {volume} {109}},\ \bibinfo {pages}
  {017201} (\bibinfo {year} {2012})}\BibitemShut {NoStop}%
\bibitem [{\citenamefont {Petit}\ \emph
  {et~al.}(2012{\natexlab{a}})\citenamefont {Petit}, \citenamefont {Bonville},
  \citenamefont {Robert}, \citenamefont {Decorse},\ and\ \citenamefont
  {Mirebeau}}]{prb.86.174403.2012}%
  \BibitemOpen
  \bibfield  {author} {\bibinfo {author} {\bibfnamefont {S.}~\bibnamefont
  {Petit}}, \bibinfo {author} {\bibfnamefont {P.}~\bibnamefont {Bonville}},
  \bibinfo {author} {\bibfnamefont {J.}~\bibnamefont {Robert}}, \bibinfo
  {author} {\bibfnamefont {C.}~\bibnamefont {Decorse}}, \ and\ \bibinfo
  {author} {\bibfnamefont {I.}~\bibnamefont {Mirebeau}},\ }\href {\doibase
  10.1103/PhysRevB.86.174403} {\bibfield  {journal} {\bibinfo  {journal} {Phys.
  Rev. B}\ }\textbf {\bibinfo {volume} {86}},\ \bibinfo {pages} {174403}
  (\bibinfo {year} {2012}{\natexlab{a}})}\BibitemShut {NoStop}%
\bibitem [{\citenamefont {Fritsch}\ \emph {et~al.}(2013)\citenamefont
  {Fritsch}, \citenamefont {Ross}, \citenamefont {Qiu}, \citenamefont {Copley},
  \citenamefont {Guidi}, \citenamefont {Bewley}, \citenamefont {Dabkowska},\
  and\ \citenamefont {Gaulin}}]{PhysRevB.87.094410}%
  \BibitemOpen
  \bibfield  {author} {\bibinfo {author} {\bibfnamefont {K.}~\bibnamefont
  {Fritsch}}, \bibinfo {author} {\bibfnamefont {K.~A.}\ \bibnamefont {Ross}},
  \bibinfo {author} {\bibfnamefont {Y.}~\bibnamefont {Qiu}}, \bibinfo {author}
  {\bibfnamefont {J.~R.~D.}\ \bibnamefont {Copley}}, \bibinfo {author}
  {\bibfnamefont {T.}~\bibnamefont {Guidi}}, \bibinfo {author} {\bibfnamefont
  {R.~I.}\ \bibnamefont {Bewley}}, \bibinfo {author} {\bibfnamefont {H.~A.}\
  \bibnamefont {Dabkowska}}, \ and\ \bibinfo {author} {\bibfnamefont {B.~D.}\
  \bibnamefont {Gaulin}},\ }\href {\doibase 10.1103/PhysRevB.87.094410}
  {\bibfield  {journal} {\bibinfo  {journal} {Phys. Rev. B}\ }\textbf {\bibinfo
  {volume} {87}},\ \bibinfo {pages} {094410} (\bibinfo {year}
  {2013})}\BibitemShut {NoStop}%
\bibitem [{\citenamefont {Molavian}\ \emph {et~al.}(2007)\citenamefont
  {Molavian}, \citenamefont {Gingras},\ and\ \citenamefont
  {Canals}}]{prl.98.157204.2007}%
  \BibitemOpen
  \bibfield  {author} {\bibinfo {author} {\bibfnamefont {H.~R.}\ \bibnamefont
  {Molavian}}, \bibinfo {author} {\bibfnamefont {M.~J.~P.}\ \bibnamefont
  {Gingras}}, \ and\ \bibinfo {author} {\bibfnamefont {B.}~\bibnamefont
  {Canals}},\ }\href@noop {} {\bibfield  {journal} {\bibinfo  {journal} {Phys.
  Rev. Lett.}\ }\textbf {\bibinfo {volume} {98}},\ \bibinfo {pages} {157204}
  (\bibinfo {year} {2007})}\BibitemShut {NoStop}%
\bibitem [{\citenamefont {Molavian}\ \emph {et~al.}(2009)\citenamefont
  {Molavian}, \citenamefont {McClarthy},\ and\ \citenamefont
  {Gingras}}]{arxiv.molavian.2009}%
  \BibitemOpen
  \bibfield  {author} {\bibinfo {author} {\bibfnamefont {H.~R.}\ \bibnamefont
  {Molavian}}, \bibinfo {author} {\bibfnamefont {P.~A.}\ \bibnamefont
  {McClarthy}}, \ and\ \bibinfo {author} {\bibfnamefont {M.~J.~P.}\
  \bibnamefont {Gingras}},\ }\href@noop {} {\enquote {\bibinfo {title}
  {{Towards an effective spin hamiltonian of the pyrochlore spin liquid
  Tb$_2$Ti$_2$O$_7$}},}\ } (\bibinfo {year} {2009}),\ \Eprint
  {http://arxiv.org/abs/0912.2957v1} {arXiv:0912.2957v1} \BibitemShut {NoStop}%
\bibitem [{\citenamefont {Mirebeau}\ \emph {et~al.}(2005)\citenamefont
  {Mirebeau}, \citenamefont {Apetrei}, \citenamefont {Rodriguez-Carvajal},
  \citenamefont {Bonville}, \citenamefont {Forget}, \citenamefont {Colson},
  \citenamefont {Glazkov}, \citenamefont {Sanchez}, \citenamefont {Isnard},\
  and\ \citenamefont {Suard}}]{PhysRevLett.94.246402}%
  \BibitemOpen
  \bibfield  {author} {\bibinfo {author} {\bibfnamefont {I.}~\bibnamefont
  {Mirebeau}}, \bibinfo {author} {\bibfnamefont {A.}~\bibnamefont {Apetrei}},
  \bibinfo {author} {\bibfnamefont {J.}~\bibnamefont {Rodriguez-Carvajal}},
  \bibinfo {author} {\bibfnamefont {P.}~\bibnamefont {Bonville}}, \bibinfo
  {author} {\bibfnamefont {A.}~\bibnamefont {Forget}}, \bibinfo {author}
  {\bibfnamefont {D.}~\bibnamefont {Colson}}, \bibinfo {author} {\bibfnamefont
  {V.}~\bibnamefont {Glazkov}}, \bibinfo {author} {\bibfnamefont {J.~P.}\
  \bibnamefont {Sanchez}}, \bibinfo {author} {\bibfnamefont {O.}~\bibnamefont
  {Isnard}}, \ and\ \bibinfo {author} {\bibfnamefont {E.}~\bibnamefont
  {Suard}},\ }\href {\doibase 10.1103/PhysRevLett.94.246402} {\bibfield
  {journal} {\bibinfo  {journal} {Phys. Rev. Lett.}\ }\textbf {\bibinfo
  {volume} {94}},\ \bibinfo {pages} {246402} (\bibinfo {year}
  {2005})}\BibitemShut {NoStop}%
\bibitem [{\citenamefont {Bonville}\ \emph {et~al.}(2011)\citenamefont
  {Bonville}, \citenamefont {Mirebeau}, \citenamefont {Gukasov}, \citenamefont
  {Petit},\ and\ \citenamefont {Robert}}]{prb.84.184409.2011}%
  \BibitemOpen
  \bibfield  {author} {\bibinfo {author} {\bibfnamefont {P.}~\bibnamefont
  {Bonville}}, \bibinfo {author} {\bibfnamefont {I.}~\bibnamefont {Mirebeau}},
  \bibinfo {author} {\bibfnamefont {A.}~\bibnamefont {Gukasov}}, \bibinfo
  {author} {\bibfnamefont {S.}~\bibnamefont {Petit}}, \ and\ \bibinfo {author}
  {\bibfnamefont {J.}~\bibnamefont {Robert}},\ }\href@noop {} {\bibfield
  {journal} {\bibinfo  {journal} {Phys. Rev. B}\ }\textbf {\bibinfo {volume}
  {84}},\ \bibinfo {pages} {184409} (\bibinfo {year} {2011})}\BibitemShut
  {NoStop}%
\bibitem [{\citenamefont {Petit}\ \emph
  {et~al.}(2012{\natexlab{b}})\citenamefont {Petit}, \citenamefont {Bonville},
  \citenamefont {Mirebeau}, \citenamefont {Mutka},\ and\ \citenamefont
  {Robert}}]{prb.85.054428.2012}%
  \BibitemOpen
  \bibfield  {author} {\bibinfo {author} {\bibfnamefont {S.}~\bibnamefont
  {Petit}}, \bibinfo {author} {\bibfnamefont {P.}~\bibnamefont {Bonville}},
  \bibinfo {author} {\bibfnamefont {I.}~\bibnamefont {Mirebeau}}, \bibinfo
  {author} {\bibfnamefont {H.}~\bibnamefont {Mutka}}, \ and\ \bibinfo {author}
  {\bibfnamefont {J.}~\bibnamefont {Robert}},\ }\href {\doibase
  10.1103/PhysRevB.85.054428} {\bibfield  {journal} {\bibinfo  {journal} {Phys.
  Rev. B}\ }\textbf {\bibinfo {volume} {85}},\ \bibinfo {pages} {054428}
  (\bibinfo {year} {2012}{\natexlab{b}})}\BibitemShut {NoStop}%
\bibitem [{\citenamefont {Ruff}\ \emph {et~al.}(2007)\citenamefont {Ruff},
  \citenamefont {Gaulin}, \citenamefont {Castellan}, \citenamefont {Rule},
  \citenamefont {Clancy}, \citenamefont {Rodriguez},\ and\ \citenamefont
  {Dabkowska}}]{prl.99.237202.2007}%
  \BibitemOpen
  \bibfield  {author} {\bibinfo {author} {\bibfnamefont {J.~P.~C.}\
  \bibnamefont {Ruff}}, \bibinfo {author} {\bibfnamefont {B.~D.}\ \bibnamefont
  {Gaulin}}, \bibinfo {author} {\bibfnamefont {J.~P.}\ \bibnamefont
  {Castellan}}, \bibinfo {author} {\bibfnamefont {K.~C.}\ \bibnamefont {Rule}},
  \bibinfo {author} {\bibfnamefont {J.~P.}\ \bibnamefont {Clancy}}, \bibinfo
  {author} {\bibfnamefont {J.}~\bibnamefont {Rodriguez}}, \ and\ \bibinfo
  {author} {\bibfnamefont {H.~A.}\ \bibnamefont {Dabkowska}},\ }\href@noop {}
  {\bibfield  {journal} {\bibinfo  {journal} {Phys. Rev. Lett.}\ }\textbf
  {\bibinfo {volume} {99}},\ \bibinfo {pages} {237202} (\bibinfo {year}
  {2007})}\BibitemShut {NoStop}%
\bibitem [{\citenamefont {Bonville}\ \emph {et~al.}(2013)\citenamefont
  {Bonville}, \citenamefont {Petit}, \citenamefont {Mirebeau}, \citenamefont
  {Robert}, \citenamefont {Lhotel},\ and\ \citenamefont
  {Paulsen}}]{arxiv.bonville.2013}%
  \BibitemOpen
  \bibfield  {author} {\bibinfo {author} {\bibfnamefont {P.}~\bibnamefont
  {Bonville}}, \bibinfo {author} {\bibfnamefont {S.}~\bibnamefont {Petit}},
  \bibinfo {author} {\bibfnamefont {I.}~\bibnamefont {Mirebeau}}, \bibinfo
  {author} {\bibfnamefont {J.}~\bibnamefont {Robert}}, \bibinfo {author}
  {\bibfnamefont {E.}~\bibnamefont {Lhotel}}, \ and\ \bibinfo {author}
  {\bibfnamefont {C.}~\bibnamefont {Paulsen}},\ }\href@noop {} {\enquote
  {\bibinfo {title} {{Magnetisation process in Er$_2$Ti$_2$O$_7$ and
  Tb$_2$Ti$_2$O$_7$ at very low temperature}},}\ } (\bibinfo {year} {2013}),\
  \Eprint {http://arxiv.org/abs/1302.6418v1} {arXiv:1302.6418v1} \BibitemShut
  {NoStop}%
\bibitem [{\citenamefont {Legl}\ \emph {et~al.}(2012)\citenamefont {Legl},
  \citenamefont {Krey}, \citenamefont {Dunsiger}, \citenamefont {Dabkowska},
  \citenamefont {Rodriguez}, \citenamefont {Luke},\ and\ \citenamefont
  {Pfleiderer}}]{prl.109.047201.2012}%
  \BibitemOpen
  \bibfield  {author} {\bibinfo {author} {\bibfnamefont {S.}~\bibnamefont
  {Legl}}, \bibinfo {author} {\bibfnamefont {C.}~\bibnamefont {Krey}}, \bibinfo
  {author} {\bibfnamefont {S.~R.}\ \bibnamefont {Dunsiger}}, \bibinfo {author}
  {\bibfnamefont {H.~A.}\ \bibnamefont {Dabkowska}}, \bibinfo {author}
  {\bibfnamefont {J.~A.}\ \bibnamefont {Rodriguez}}, \bibinfo {author}
  {\bibfnamefont {G.~M.}\ \bibnamefont {Luke}}, \ and\ \bibinfo {author}
  {\bibfnamefont {C.}~\bibnamefont {Pfleiderer}},\ }\href {\doibase
  10.1103/PhysRevLett.109.047201} {\bibfield  {journal} {\bibinfo  {journal}
  {Phys. Rev. Lett.}\ }\textbf {\bibinfo {volume} {109}},\ \bibinfo {pages}
  {047201} (\bibinfo {year} {2012})}\BibitemShut {NoStop}%
\bibitem [{\citenamefont {Lhotel}\ \emph {et~al.}(2012)\citenamefont {Lhotel},
  \citenamefont {Paulsen}, \citenamefont {de~R\'{e}otier}, \citenamefont
  {Yaouanc}, \citenamefont {Marin},\ and\ \citenamefont
  {Vanishri}}]{prb.86.020410.2012}%
  \BibitemOpen
  \bibfield  {author} {\bibinfo {author} {\bibfnamefont {E.}~\bibnamefont
  {Lhotel}}, \bibinfo {author} {\bibfnamefont {C.}~\bibnamefont {Paulsen}},
  \bibinfo {author} {\bibfnamefont {P.~D.}\ \bibnamefont {de~R\'{e}otier}},
  \bibinfo {author} {\bibfnamefont {A.}~\bibnamefont {Yaouanc}}, \bibinfo
  {author} {\bibfnamefont {C.}~\bibnamefont {Marin}}, \ and\ \bibinfo {author}
  {\bibfnamefont {S.}~\bibnamefont {Vanishri}},\ }\href@noop {} {\bibfield
  {journal} {\bibinfo  {journal} {Phys. Rev. B}\ }\textbf {\bibinfo {volume}
  {86}},\ \bibinfo {pages} {020410} (\bibinfo {year} {2012})}\BibitemShut
  {NoStop}%
\bibitem [{\citenamefont {Baker}\ \emph {et~al.}(2012)\citenamefont {Baker},
  \citenamefont {Matthews}, \citenamefont {Giblin}, \citenamefont {Schiffer},
  \citenamefont {Baines},\ and\ \citenamefont
  {Prabhakaran}}]{prb.86.094424.2012}%
  \BibitemOpen
  \bibfield  {author} {\bibinfo {author} {\bibfnamefont {P.~J.}\ \bibnamefont
  {Baker}}, \bibinfo {author} {\bibfnamefont {M.~J.}\ \bibnamefont {Matthews}},
  \bibinfo {author} {\bibfnamefont {S.~R.}\ \bibnamefont {Giblin}}, \bibinfo
  {author} {\bibfnamefont {P.}~\bibnamefont {Schiffer}}, \bibinfo {author}
  {\bibfnamefont {C.}~\bibnamefont {Baines}}, \ and\ \bibinfo {author}
  {\bibfnamefont {D.}~\bibnamefont {Prabhakaran}},\ }\href {\doibase
  10.1103/PhysRevB.86.094424} {\bibfield  {journal} {\bibinfo  {journal} {Phys.
  Rev. B}\ }\textbf {\bibinfo {volume} {86}},\ \bibinfo {pages} {094424}
  (\bibinfo {year} {2012})}\BibitemShut {NoStop}%
\bibitem [{\citenamefont {Yin}\ \emph {et~al.}(2013)\citenamefont {Yin},
  \citenamefont {Xia}, \citenamefont {Takano}, \citenamefont {Sullivan},
  \citenamefont {Li},\ and\ \citenamefont {Sun}}]{PhysRevLett.110.137201}%
  \BibitemOpen
  \bibfield  {author} {\bibinfo {author} {\bibfnamefont {L.}~\bibnamefont
  {Yin}}, \bibinfo {author} {\bibfnamefont {J.~S.}\ \bibnamefont {Xia}},
  \bibinfo {author} {\bibfnamefont {Y.}~\bibnamefont {Takano}}, \bibinfo
  {author} {\bibfnamefont {N.~S.}\ \bibnamefont {Sullivan}}, \bibinfo {author}
  {\bibfnamefont {Q.~J.}\ \bibnamefont {Li}}, \ and\ \bibinfo {author}
  {\bibfnamefont {X.~F.}\ \bibnamefont {Sun}},\ }\href {\doibase
  10.1103/PhysRevLett.110.137201} {\bibfield  {journal} {\bibinfo  {journal}
  {Phys. Rev. Lett.}\ }\textbf {\bibinfo {volume} {110}},\ \bibinfo {pages}
  {137201} (\bibinfo {year} {2013})}\BibitemShut {NoStop}%
\bibitem [{\citenamefont {Chapuis}\ \emph {et~al.}(2010)\citenamefont
  {Chapuis}, \citenamefont {Yaouanc}, \citenamefont {Dalmas~de R\'eotier},
  \citenamefont {Marin}, \citenamefont {Vanishri}, \citenamefont {Curnoe},
  \citenamefont {V\^aju},\ and\ \citenamefont {Forget}}]{prb.82.100402.2010}%
  \BibitemOpen
  \bibfield  {author} {\bibinfo {author} {\bibfnamefont {Y.}~\bibnamefont
  {Chapuis}}, \bibinfo {author} {\bibfnamefont {A.}~\bibnamefont {Yaouanc}},
  \bibinfo {author} {\bibfnamefont {P.}~\bibnamefont {Dalmas~de R\'eotier}},
  \bibinfo {author} {\bibfnamefont {C.}~\bibnamefont {Marin}}, \bibinfo
  {author} {\bibfnamefont {S.}~\bibnamefont {Vanishri}}, \bibinfo {author}
  {\bibfnamefont {S.~H.}\ \bibnamefont {Curnoe}}, \bibinfo {author}
  {\bibfnamefont {C.}~\bibnamefont {V\^aju}}, \ and\ \bibinfo {author}
  {\bibfnamefont {A.}~\bibnamefont {Forget}},\ }\href {\doibase
  10.1103/PhysRevB.82.100402} {\bibfield  {journal} {\bibinfo  {journal} {Phys.
  Rev. B}\ }\textbf {\bibinfo {volume} {82}},\ \bibinfo {pages} {100402}
  (\bibinfo {year} {2010})}\BibitemShut {NoStop}%
\bibitem [{\citenamefont {Rule}\ \emph {et~al.}(2006)\citenamefont {Rule},
  \citenamefont {Ruff}, \citenamefont {Gaulin}, \citenamefont {Dunsiger},
  \citenamefont {Gardner}, \citenamefont {Clancy}, \citenamefont {Lewis},
  \citenamefont {Dabkowska}, \citenamefont {Mirebeau}, \citenamefont {Manuel},
  \citenamefont {Qiu},\ and\ \citenamefont {Copley}}]{prl.96.177201.2006}%
  \BibitemOpen
  \bibfield  {author} {\bibinfo {author} {\bibfnamefont {K.~C.}\ \bibnamefont
  {Rule}}, \bibinfo {author} {\bibfnamefont {J.~P.~C.}\ \bibnamefont {Ruff}},
  \bibinfo {author} {\bibfnamefont {B.~D.}\ \bibnamefont {Gaulin}}, \bibinfo
  {author} {\bibfnamefont {S.~R.}\ \bibnamefont {Dunsiger}}, \bibinfo {author}
  {\bibfnamefont {J.~S.}\ \bibnamefont {Gardner}}, \bibinfo {author}
  {\bibfnamefont {J.~P.}\ \bibnamefont {Clancy}}, \bibinfo {author}
  {\bibfnamefont {M.~J.}\ \bibnamefont {Lewis}}, \bibinfo {author}
  {\bibfnamefont {H.~A.}\ \bibnamefont {Dabkowska}}, \bibinfo {author}
  {\bibfnamefont {I.}~\bibnamefont {Mirebeau}}, \bibinfo {author}
  {\bibfnamefont {P.}~\bibnamefont {Manuel}}, \bibinfo {author} {\bibfnamefont
  {Y.}~\bibnamefont {Qiu}}, \ and\ \bibinfo {author} {\bibfnamefont {J.~R.~D.}\
  \bibnamefont {Copley}},\ }\href@noop {} {\bibfield  {journal} {\bibinfo
  {journal} {Phys. Rev. Lett.}\ }\textbf {\bibinfo {volume} {96}},\ \bibinfo
  {pages} {177201} (\bibinfo {year} {2006})}\BibitemShut {NoStop}%
\bibitem [{\citenamefont {Cao}\ \emph {et~al.}(2008)\citenamefont {Cao},
  \citenamefont {Gukasov}, \citenamefont {Mirebeau}, \citenamefont {Bonville},\
  and\ \citenamefont {Dhalenne}}]{prl.101.196402.2008}%
  \BibitemOpen
  \bibfield  {author} {\bibinfo {author} {\bibfnamefont {H.}~\bibnamefont
  {Cao}}, \bibinfo {author} {\bibfnamefont {A.}~\bibnamefont {Gukasov}},
  \bibinfo {author} {\bibfnamefont {I.}~\bibnamefont {Mirebeau}}, \bibinfo
  {author} {\bibfnamefont {P.}~\bibnamefont {Bonville}}, \ and\ \bibinfo
  {author} {\bibfnamefont {G.}~\bibnamefont {Dhalenne}},\ }\href@noop {}
  {\bibfield  {journal} {\bibinfo  {journal} {Phys. Rev. Lett.}\ }\textbf
  {\bibinfo {volume} {101}},\ \bibinfo {pages} {196402} (\bibinfo {year}
  {2008})}\BibitemShut {NoStop}%
\bibitem [{\citenamefont {Sazonov}\ \emph {et~al.}(2010)\citenamefont
  {Sazonov}, \citenamefont {Gukasov}, \citenamefont {Mirebeau}, \citenamefont
  {Cao}, \citenamefont {Bonville}, \citenamefont {Grenier},\ and\ \citenamefont
  {Dhalenne}}]{prb.82.174406.2010}%
  \BibitemOpen
  \bibfield  {author} {\bibinfo {author} {\bibfnamefont {A.~P.}\ \bibnamefont
  {Sazonov}}, \bibinfo {author} {\bibfnamefont {A.}~\bibnamefont {Gukasov}},
  \bibinfo {author} {\bibfnamefont {I.}~\bibnamefont {Mirebeau}}, \bibinfo
  {author} {\bibfnamefont {H.}~\bibnamefont {Cao}}, \bibinfo {author}
  {\bibfnamefont {P.}~\bibnamefont {Bonville}}, \bibinfo {author}
  {\bibfnamefont {B.}~\bibnamefont {Grenier}}, \ and\ \bibinfo {author}
  {\bibfnamefont {G.}~\bibnamefont {Dhalenne}},\ }\href@noop {} {\bibfield
  {journal} {\bibinfo  {journal} {Phys. Rev. B}\ }\textbf {\bibinfo {volume}
  {82}},\ \bibinfo {pages} {174406} (\bibinfo {year} {2010})}\BibitemShut
  {NoStop}%
\bibitem [{\citenamefont {Sazonov}\ \emph {et~al.}(2012)\citenamefont
  {Sazonov}, \citenamefont {Gukasov}, \citenamefont {Mirebeau},\ and\
  \citenamefont {Bonville}}]{prb.85.214420.2012}%
  \BibitemOpen
  \bibfield  {author} {\bibinfo {author} {\bibfnamefont {A.~P.}\ \bibnamefont
  {Sazonov}}, \bibinfo {author} {\bibfnamefont {A.}~\bibnamefont {Gukasov}},
  \bibinfo {author} {\bibfnamefont {I.}~\bibnamefont {Mirebeau}}, \ and\
  \bibinfo {author} {\bibfnamefont {P.}~\bibnamefont {Bonville}},\ }\href
  {\doibase 10.1103/PhysRevB.85.214420} {\bibfield  {journal} {\bibinfo
  {journal} {Phys. Rev. B}\ }\textbf {\bibinfo {volume} {85}},\ \bibinfo
  {pages} {214420} (\bibinfo {year} {2012})}\BibitemShut {NoStop}%
\bibitem [{\citenamefont {Yasui}\ \emph {et~al.}(2001)\citenamefont {Yasui},
  \citenamefont {Kanada}, \citenamefont {Ito}, \citenamefont {Harashina},
  \citenamefont {Sato}, \citenamefont {Okumura},\ and\ \citenamefont
  {Kakurai}}]{jpcs.62.343.2001}%
  \BibitemOpen
  \bibfield  {author} {\bibinfo {author} {\bibfnamefont {Y.}~\bibnamefont
  {Yasui}}, \bibinfo {author} {\bibfnamefont {M.}~\bibnamefont {Kanada}},
  \bibinfo {author} {\bibfnamefont {M.}~\bibnamefont {Ito}}, \bibinfo {author}
  {\bibfnamefont {H.}~\bibnamefont {Harashina}}, \bibinfo {author}
  {\bibfnamefont {M.}~\bibnamefont {Sato}}, \bibinfo {author} {\bibfnamefont
  {H.}~\bibnamefont {Okumura}}, \ and\ \bibinfo {author} {\bibfnamefont
  {K.}~\bibnamefont {Kakurai}},\ }\href@noop {} {\bibfield  {journal} {\bibinfo
   {journal} {J. Phys. Chem. Solids}\ }\textbf {\bibinfo {volume} {62}},\
  \bibinfo {pages} {343} (\bibinfo {year} {2001})}\BibitemShut {NoStop}%
\bibitem [{\citenamefont {Rodr{\'\i}guez-Carvajal}(1993)}]{phb.192.55.1993}%
  \BibitemOpen
  \bibfield  {author} {\bibinfo {author} {\bibfnamefont {J.}~\bibnamefont
  {Rodr{\'\i}guez-Carvajal}},\ }\href@noop {} {\bibfield  {journal} {\bibinfo
  {journal} {Physica B}\ }\textbf {\bibinfo {volume} {192}},\ \bibinfo {pages}
  {55} (\bibinfo {year} {1993})}\BibitemShut {NoStop}%
\bibitem [{\citenamefont {Bertaut}(1968)}]{acra.24.217.1968}%
  \BibitemOpen
  \bibfield  {author} {\bibinfo {author} {\bibfnamefont {E.~F.}\ \bibnamefont
  {Bertaut}},\ }\href@noop {} {\bibfield  {journal} {\bibinfo  {journal} {Acta
  Cryst. A}\ }\textbf {\bibinfo {volume} {24}},\ \bibinfo {pages} {217}
  (\bibinfo {year} {1968})}\BibitemShut {NoStop}%
\bibitem [{\citenamefont {Izyumov}\ and\ \citenamefont
  {Naish}(1979)}]{jmmm.12.239.1979}%
  \BibitemOpen
  \bibfield  {author} {\bibinfo {author} {\bibfnamefont {Y.~A.}\ \bibnamefont
  {Izyumov}}\ and\ \bibinfo {author} {\bibfnamefont {V.~E.}\ \bibnamefont
  {Naish}},\ }\href@noop {} {\bibfield  {journal} {\bibinfo  {journal} {J.
  Magn. Magn. Mater.}\ }\textbf {\bibinfo {volume} {12}},\ \bibinfo {pages}
  {239} (\bibinfo {year} {1979})}\BibitemShut {NoStop}%
\bibitem [{\citenamefont {Kovalev}(1965)}]{book.kovalev.1965}%
  \BibitemOpen
  \bibfield  {author} {\bibinfo {author} {\bibfnamefont {O.~V.}\ \bibnamefont
  {Kovalev}},\ }\href@noop {} {\emph {\bibinfo {title} {Irreducible
  Representations of the Space Groups}}}\ (\bibinfo  {publisher} {Gordon and
  Breach, New York},\ \bibinfo {year} {1965})\BibitemShut {NoStop}%
\bibitem [{\citenamefont {Rule}\ and\ \citenamefont
  {Bonville}(2009)}]{jpcs.145.012027.2009}%
  \BibitemOpen
  \bibfield  {author} {\bibinfo {author} {\bibfnamefont {K.~C.}\ \bibnamefont
  {Rule}}\ and\ \bibinfo {author} {\bibfnamefont {P.}~\bibnamefont
  {Bonville}},\ }\href@noop {} {\bibfield  {journal} {\bibinfo  {journal} {J.
  Phys.: Conf. Ser.}\ }\textbf {\bibinfo {volume} {145}},\ \bibinfo {pages}
  {012027} (\bibinfo {year} {2009})}\BibitemShut {NoStop}%
\bibitem [{\citenamefont {Tariq}\ \emph {et~al.}(2013)\citenamefont {Tariq},
  \citenamefont {Taisan}, \citenamefont {Singh},\ and\ \citenamefont
  {Weinstein}}]{PhysRevLett.110.153201}%
  \BibitemOpen
  \bibfield  {author} {\bibinfo {author} {\bibfnamefont {N.}~\bibnamefont
  {Tariq}}, \bibinfo {author} {\bibfnamefont {N.~A.}\ \bibnamefont {Taisan}},
  \bibinfo {author} {\bibfnamefont {V.}~\bibnamefont {Singh}}, \ and\ \bibinfo
  {author} {\bibfnamefont {J.~D.}\ \bibnamefont {Weinstein}},\ }\href {\doibase
  10.1103/PhysRevLett.110.153201} {\bibfield  {journal} {\bibinfo  {journal}
  {Phys. Rev. Lett.}\ }\textbf {\bibinfo {volume} {110}},\ \bibinfo {pages}
  {153201} (\bibinfo {year} {2013})}\BibitemShut {NoStop}%
\bibitem [{\citenamefont {Mamsurova}\ \emph {et~al.}(1986)\citenamefont
  {Mamsurova}, \citenamefont {Pigal'skii},\ and\ \citenamefont
  {Pukhov}}]{mamsurova}%
  \BibitemOpen
  \bibfield  {author} {\bibinfo {author} {\bibfnamefont {L.~G.}\ \bibnamefont
  {Mamsurova}}, \bibinfo {author} {\bibfnamefont {K.~S.}\ \bibnamefont
  {Pigal'skii}}, \ and\ \bibinfo {author} {\bibfnamefont {K.~K.}\ \bibnamefont
  {Pukhov}},\ }\href@noop {} {\bibfield  {journal} {\bibinfo  {journal} {JETP
  Letters}\ }\textbf {\bibinfo {volume} {43}},\ \bibinfo {pages} {755}
  (\bibinfo {year} {1986})}\BibitemShut {NoStop}%
\bibitem [{\citenamefont {Nakanishi}\ \emph {et~al.}(2011)\citenamefont
  {Nakanishi}, \citenamefont {Kumagai}, \citenamefont {Yoshizawa},
  \citenamefont {Matsuhira}, \citenamefont {Takagi},\ and\ \citenamefont
  {Hiroi}}]{nakanishi}%
  \BibitemOpen
  \bibfield  {author} {\bibinfo {author} {\bibfnamefont {Y.}~\bibnamefont
  {Nakanishi}}, \bibinfo {author} {\bibfnamefont {T.}~\bibnamefont {Kumagai}},
  \bibinfo {author} {\bibfnamefont {M.}~\bibnamefont {Yoshizawa}}, \bibinfo
  {author} {\bibfnamefont {K.}~\bibnamefont {Matsuhira}}, \bibinfo {author}
  {\bibfnamefont {S.}~\bibnamefont {Takagi}}, \ and\ \bibinfo {author}
  {\bibfnamefont {Z.}~\bibnamefont {Hiroi}},\ }\href@noop {} {\bibfield
  {journal} {\bibinfo  {journal} {Phys. Rev. B}\ }\textbf {\bibinfo {volume}
  {83}},\ \bibinfo {pages} {184434} (\bibinfo {year} {2011})}\BibitemShut
  {NoStop}%
\bibitem [{\citenamefont {Luan}(2011)}]{luan}%
  \BibitemOpen
  \bibfield  {author} {\bibinfo {author} {\bibfnamefont {Y.}~\bibnamefont
  {Luan}},\ }\href {{http://trace.tennessee.edu/utk_graddiss/993}} {\enquote
  {\bibinfo {title} {{Elastic properties of complex transition metal oxides
  studied by Resonant Ultrasound Spectroscopy}},}\ } (\bibinfo {year}
  {2011})\BibitemShut {NoStop}%
\bibitem [{\citenamefont {Klekovkina}\ \emph {et~al.}(2011)\citenamefont
  {Klekovkina}, \citenamefont {Zakirov}, \citenamefont {Malkin},\ and\
  \citenamefont {Kasatkina}}]{jpcs.324.012036.2011}%
  \BibitemOpen
  \bibfield  {author} {\bibinfo {author} {\bibfnamefont {V.~V.}\ \bibnamefont
  {Klekovkina}}, \bibinfo {author} {\bibfnamefont {A.~R.}\ \bibnamefont
  {Zakirov}}, \bibinfo {author} {\bibfnamefont {B.~Z.}\ \bibnamefont {Malkin}},
  \ and\ \bibinfo {author} {\bibfnamefont {L.~A.}\ \bibnamefont {Kasatkina}},\
  }\href@noop {} {\bibfield  {journal} {\bibinfo  {journal} {J. Phys.: Conf.
  Ser.}\ }\textbf {\bibinfo {volume} {324}},\ \bibinfo {pages} {012036}
  (\bibinfo {year} {2011})}\BibitemShut {NoStop}%
\bibitem [{\citenamefont {Guitteny}\ \emph {et~al.}(2013)\citenamefont
  {Guitteny}, \citenamefont {Robert}, \citenamefont {Bonville}, \citenamefont
  {Decorse}, \citenamefont {Steffens}, \citenamefont {Boehm}, \citenamefont
  {Mutka}, \citenamefont {Ollivier}, \citenamefont {Mirebeau},\ and\
  \citenamefont {Petit}}]{arxiv.guitteny.2013}%
  \BibitemOpen
  \bibfield  {author} {\bibinfo {author} {\bibfnamefont {S.}~\bibnamefont
  {Guitteny}}, \bibinfo {author} {\bibfnamefont {J.}~\bibnamefont {Robert}},
  \bibinfo {author} {\bibfnamefont {P.}~\bibnamefont {Bonville}}, \bibinfo
  {author} {\bibfnamefont {C.}~\bibnamefont {Decorse}}, \bibinfo {author}
  {\bibfnamefont {P.}~\bibnamefont {Steffens}}, \bibinfo {author}
  {\bibfnamefont {M.}~\bibnamefont {Boehm}}, \bibinfo {author} {\bibfnamefont
  {H.}~\bibnamefont {Mutka}}, \bibinfo {author} {\bibfnamefont
  {J.}~\bibnamefont {Ollivier}}, \bibinfo {author} {\bibfnamefont
  {I.}~\bibnamefont {Mirebeau}}, \ and\ \bibinfo {author} {\bibfnamefont
  {S.}~\bibnamefont {Petit}},\ }\href@noop {} {\enquote {\bibinfo {title}
  {{Anisotropic propagative excitations and quadrupolar effects in
  Tb$_2$Ti$_2$O$_7$}},}\ } (\bibinfo {year} {2013}),\ \Eprint
  {http://arxiv.org/abs/1305.6363v1} {arXiv:1305.6363v1} \BibitemShut {NoStop}%
\bibitem [{\citenamefont {Fennell}\ \emph {et~al.}(2013)\citenamefont
  {Fennell}, \citenamefont {Kenzelmann}, \citenamefont {Roessli}, \citenamefont
  {Mutka}, \citenamefont {Ollivier}, \citenamefont {Ruminy}, \citenamefont
  {Stuhr}, \citenamefont {Zaharko}, \citenamefont {Bovo}, \citenamefont
  {Cervellino}, \citenamefont {Haas},\ and\ \citenamefont
  {Cava}}]{arxiv.fennell.2013}%
  \BibitemOpen
  \bibfield  {author} {\bibinfo {author} {\bibfnamefont {T.}~\bibnamefont
  {Fennell}}, \bibinfo {author} {\bibfnamefont {M.}~\bibnamefont {Kenzelmann}},
  \bibinfo {author} {\bibfnamefont {B.}~\bibnamefont {Roessli}}, \bibinfo
  {author} {\bibfnamefont {H.}~\bibnamefont {Mutka}}, \bibinfo {author}
  {\bibfnamefont {J.}~\bibnamefont {Ollivier}}, \bibinfo {author}
  {\bibfnamefont {M.}~\bibnamefont {Ruminy}}, \bibinfo {author} {\bibfnamefont
  {U.}~\bibnamefont {Stuhr}}, \bibinfo {author} {\bibfnamefont
  {O.}~\bibnamefont {Zaharko}}, \bibinfo {author} {\bibfnamefont
  {L.}~\bibnamefont {Bovo}}, \bibinfo {author} {\bibfnamefont {A.}~\bibnamefont
  {Cervellino}}, \bibinfo {author} {\bibfnamefont {M.~K.}\ \bibnamefont
  {Haas}}, \ and\ \bibinfo {author} {\bibfnamefont {R.~J.}\ \bibnamefont
  {Cava}},\ }\href@noop {} {\enquote {\bibinfo {title} {{Magnetoelastic
  excitations in the pyrochlore spin liquid Tb$_2$Ti$_2$O$_7$}},}\ } (\bibinfo
  {year} {2013}),\ \Eprint {http://arxiv.org/abs/1305.5405v1}
  {arXiv:1305.5405v1} \BibitemShut {NoStop}%
\bibitem [{\citenamefont {Li}\ \emph {et~al.}(2013)\citenamefont {Li},
  \citenamefont {Zhao}, \citenamefont {Fan}, \citenamefont {Zhang},
  \citenamefont {Zhou}, \citenamefont {Zhao},\ and\ \citenamefont
  {Sun}}]{arxiv.li.2013}%
  \BibitemOpen
  \bibfield  {author} {\bibinfo {author} {\bibfnamefont {Q.~J.}\ \bibnamefont
  {Li}}, \bibinfo {author} {\bibfnamefont {Z.~Y.}\ \bibnamefont {Zhao}},
  \bibinfo {author} {\bibfnamefont {C.}~\bibnamefont {Fan}}, \bibinfo {author}
  {\bibfnamefont {F.~B.}\ \bibnamefont {Zhang}}, \bibinfo {author}
  {\bibfnamefont {H.~D.}\ \bibnamefont {Zhou}}, \bibinfo {author}
  {\bibfnamefont {X.}~\bibnamefont {Zhao}}, \ and\ \bibinfo {author}
  {\bibfnamefont {X.~F.}\ \bibnamefont {Sun}},\ }\href@noop {} {\enquote
  {\bibinfo {title} {{Magnetically originated phonon-glass-like behavior in
  Tb$_2$Ti$_2$O$_7$ single crystal}},}\ } (\bibinfo {year} {2013}),\ \Eprint
  {http://arxiv.org/abs/1305.6853v1} {arXiv:1305.6853v1} \BibitemShut {NoStop}%
\end{thebibliography}



%

\end{document}